\documentclass[12pt,reprint,tightenlines,groupedaddress,superscriptaddress,floatfix,aps]{revtex4}
\usepackage{graphicx}
\usepackage{rotating}
\usepackage{array}
\usepackage{multirow}
\usepackage{epstopdf}
\usepackage{amsmath}
\usepackage{amsfonts}
\usepackage{setspace}
\usepackage{braket}
\usepackage{color}

\usepackage{hyperref}
\usepackage[T1]{fontenc}

\DeclareMathOperator{\Tr}{Tr}

\begin{document}
\title{Natural orbitals and two-particle correlators as tools for analysis of effective exchange couplings in solids} 

\author{Pavel Pokhilko}
\affiliation{Department  of  Chemistry,  University  of  Michigan,  Ann  Arbor,  Michigan  48109,  USA}
\author{Dominika Zgid}
\affiliation{Department  of  Chemistry,  University  of  Michigan,  Ann  Arbor,  Michigan  48109,  USA}
\affiliation{Department of Physics, University of Michigan, Ann Arbor, Michigan 48109, USA }

\renewcommand{\baselinestretch}{1.0}
\begin{abstract}
Using generalizations of natural orbitals, spin-averaged natural orbitals, and two-particle charge correlators for solids, 
we investigate electronic structure of antiferromagnetic transition-metal oxides 
with a fully self-consistent, finite-temperature GW method.  
Our findings disagree with Goodenough--Kanamori (GK) rules, commonly used for qualitative interpretation of such solids. 
First, we found a strong dependence of natural orbital occupancies on momenta contradicting GK assumptions. 
Second, along the momentum path, the character of natural orbitals changes. 
In particular, contributions of oxygen $2s$ orbitals are important, which has not been considered in the GK rules. 
To analyze the influence of electronic correlation on the values of effective exchange coupling constants, we use both natural orbitals and two-particle correlators and show that electronic screening modulates 
the degree of superexchange by stabilizing the charge-transfer contributions, greatly affecting these coupling constants. 
Finally, we give a set of predictions and recommendations regarding the use of density functional, Green's function, and wave-function methods for evaluating effective magnetic couplings in molecules and solids. 
\end{abstract}
\maketitle

\section{Introduction}

The wave function employed in the description of magnetic phenomena requires multiple configurations to quantitatively and qualitatively describe magnetic states.
Since the treatment of both strong and weak electron correlation is necessary, a number of strategies for evaluation of molecular magnetic couplings has been used in the past.
Among the active-space wave-function methods, 
various selective and multireference configuration interaction\cite{Malrieu:DDCI:1993,Zimmerman:iFCI:exchange:2021} and 
perturbation theories\cite{Roos:90:CASPT2,Angeli1,Angeli2} are commonly used for 
small molecules where such calculations are affordable.  
While strong correlation is important for the description of magnetic interactions, 
explicit multireference methods are not the only family of approaches that can accurately capture it. 
For example, magnetic couplings can also be efficiently and accurately computed within 
spin-flip family of methods\cite{Casanova:SFReview} 
that includes several spin-flip time-dependent density functional theory (SF-TDDFT) formulations\cite{Krylov:SFTDDFT:2003,Bernard:SF:12,Ziegler:sfdft:04,Ziegler:sa:cvdft:11,Ziegler:Cu2:2011,Ziegler:Cu2tb:2012,Ziegler:Cu3:11,Liu:SFtensor:10,Liu:SFtensor:11,Valero:SFDFT:11,Orms:magnets:17,Kotaru:Fe:SMM:2022} 
and equation-of-motion coupled-cluster methods\cite{sfpaper,Casanova:2SF:08} (EOM-SF-CC) or its perturbative approximations\cite{Jagau:SF-CC2:2022}.  
Broken-symmetry approaches form another family of methods that when combined with DFT (denoted as BS-DFT) are the most 
common tools for evaluating magnetic couplings due to their low computational cost.
Broken-symmetry approaches have also been deployed within a number of wave-function\cite{Yamaguchi:APUMP:1989,Yamaguchi:APCCSD:2012,Stanton:BS-CC:2020} and Green's function methods\cite{Chibotaru:BS-G0W0:2020,Pokhilko:local_correlators:2021,Pokhilko:BS-GW:solids:2022}. Another approach is to use a magnetic force theorem\cite{Liechtenstein:mag_force:1987}, which is even cheaper than BS-DFT when deployed within DFT\cite{Peralta:mag_force:2022}.  

For solids, due to high computational cost and necessity of reaching solutions in thermodynamic limit (to remove finite-size effects), 
the evaluation of magnetic couplings within the wave-function formalism was rarely performed before\cite{Martin:NiO:exchange:2002,deGraaf:MRPT:exchange:solids:SMM:2001,Pokhilko:spinchain}. 
The majority of magnetic exchange calculations in solids was performed using BS-DFT. 
However, BS-DFT has numerous disadvantages. 
First, it depends on a particular functional parametrization (in the Kohn--Sham formulation); thus, it is not systematically improvable. 
Second, the extraction procedure of magnetic couplings from the BS-DFT solutions is not well defined and a number of approximate spin-projection and decontamination schemes have been employed\cite{noodleman:BS:81,Yamaguchi:BS:formulation:1986,Malrieu:spin_pol:BS-DFT:2020,Malrieu:decont:BS-DFT:2020} in the past. 
Moreover, $\braket{S^2}$ needed for such spin projections often is not evaluated rigorously within DFT\cite{Cremer:DFT:S2:2001,Handy:DFT:S2:2007,Vedene:DFT:S2:1995}, 
which ultimately introduces additional errors. 
Analysis of spin contamination in solids is even more complicated and is based on various spin--spin correlation functions\cite{Yamaguchi:APDFT:solid:2021,Pokhilko:BS-GW:solids:2022}.

Green's function methods\cite{Mahan00,Negele:Orland:book:2018,Martin:Interacting_electrons:2016} offer a viable and affordable alternative for treating magnetic phenomena in solids.
On one hand, unlike bulky multi-index wave-function amplitudes, 
the one-electron Green's function depends only on two orbital and one time index resulting in a   very compact object suitable for the development of computationally efficient methods. 
On the other hand the Green's function and self-energy contain a lot of information about the system 
and can be used to evaluate one- and some of the two-particle observables directly. 
Due to these features, these powerful methods found numerous applications in, for instance, 
in photoelectron spectroscopy\cite{Almbladh:photoemission:1985,Hedin:photoemission:1985,Fujikawa:photoelectron:chapter:2015} and superconductivity\cite{Kadanoff:superconductivity:1961}.  

Fully self-consistent Green's function methods do not depend on any adjustable parameters 
and have a systematically improvable perturbative expansion given by skeleton (bold) diagrams\cite{Luttinger60}. 
The Green's function methods based on such a bold diagrammatic expansion are also called conserving, 
since for them the Kadamoff--Baym theorem\cite{Baym61,Baym62} grants the conservation of 
momentum, angular momentum, energy as well as gauge invariance of the Luttinger--Ward functional and the current continuity. 
Full self-consistency removes any dependence on a starting point, 
which is not removed within the non-self-consistent and partly self-consistent methods. 
Recently, we have introduced an $\braket{S^2}$ evaluation in  
the self-consistent Green's function formalism\cite{Pokhilko:tpdm:2021} 
and found that in many cases the degree of spin contamination between different magnetic solutions in molecules\cite{Pokhilko:local_correlators:2021} and solids\cite{Pokhilko:BS-GW:solids:2022} is very small, making the extraction procedure rigorous, unambiguous, and simple. 

Effective Hamitonian theory\cite{Cloizeax:1960,Bloch:1958,Okubo:1954,Durand:EffHam:1983,Soliverez:1969} offers a straightforward way for constructing magnetic Hamiltonians. This theoretical framework has been used within a number of methods\cite{Calzado:02,Marlieu:MagnetRev:2014} both for the derivation purposes as well as for constructing new approximations based on the physics of the effective Hamiltonians\cite{Buchachenko:Mn2:2010,Mayhall:2014:HDVV,Mayhall:1SF:2015,Pokhilko:EffH:2020,Pokhilko:Neel_T:2022}.  
Another but related strategy utilizes finite-order perturbative expansions of effective Hamiltonians and investigates configurations contributing to the effective exchange coupling. 
For example, Kramers\cite{Kramers:superexchange:1934} and Anderson\cite{Anderson:superexchange:1950} 
used the latter strategy, and classified various terms 
into direct exchange, kinetic exchange, and spin-polarisation terms\cite{Anderson:exchange:1963,Eremin:exchange:1980}.  
Unfortunately, such terminology is often misleading since these terms are defined differently in various publications. 
Such an analysis (sometimes emphasized as an exchange decomposition) has been less rigorously extended to BS-DFT\cite{Coulaud:Jdecomp:2012,Coulaud:Jdecomp:2013,Ferre:Jdecomp:2018,LeGuennic:multicenter:Jdecomp:2023}, 
where such terms are defined through ``energies'' of various unoptimized, partly optimized, and fully optimized restricted open-shell determinants with approximate spin projections. 
Besides ambiguity, an apparent limitation of this decomposition is in its strong dependence on a particular family of methods, making it not applicable for the analysis and comparisons of magnetic couplings between various electronic structure methods.  

A robust analysis shall probe the nature of the system \emph{directly} and without any dependence on a particular flavor of electronic structure method. For example, natural orbitals and two-particle correlators provide such an ansatz-agnostic analysis, which made them applicable for both wave-function and Green's function methods\cite{Pokhilko:local_correlators:2021}. 
This unified language not only gives an insight into basic physics of the particular system, but also allows one to look at electronic structure from different perspectives and to establish deeper connections between different families of methods.  

In this paper, we extend the analysis of molecular electronic structure based on natural orbitals and 
two-particle charge correlators to solids and deploy it within Green's function methods. 
Such a generalization allows us to quantify and analyze the location of open-shell electrons, 
the degree of open-shell and charge-transfer nature, 
and the role of electron correlation in transition-metal oxides---systems with a substantial presence of superexchange. 
Based on this analysis, we gain a deeper insight into the physics of magnetic interactions and 
formulate physical predictions that can be checked numerically in the future. 

\section{Theory}
\subsection{Green's functions and density matrices}
We choose to work with Bloch orbitals since they allow us to directly \emph{translate} electronic structure methods 
from molecules to solids. 
In the same way, one can also transfer the analysis of electronic structure, which we pursue in this paper. 
The Bloch orbitals are defined as
\begin{gather}
\phi_{\mathbf{k},i} = \sum_{\mathbf{R}} \phi_i^{\mathbf{R}} (\mathbf{r}) e^{i\mathbf{k}\cdot\mathbf{R}},
\protect\label{eq:Bloch}
\end{gather}
where $\mathbf{k}$ is the momentum vector index, 
$i$ subscript enumerates atom-centered combination of primitive Gaussians $\phi_i^{\mathbf{R}} (\mathbf{r})$ in the unit cell,
$\mathbf{R}$ is a Bravais lattice position, $\mathbf{r}$ is the space electronic coordinate. 
For convenience, we merge spin, momentum, and space coordinates\cite{note:time} and 
label the corresponding momentum-dependent spin-orbitals with multi-indices: $\mathbf{p,q,r,s}$. 
Such notation is very compact and analogous to the molecular case, making generalizations to solids transparent. 
For example, one-particle Green's functions\cite{Mahan00,Negele:Orland:book:2018,Martin:Interacting_electrons:2016} are defined as
\begin{gather}
G_{\mathbf{pq}} (\tau) = -\frac{1}{Z} \Tr \left[e^{-(\beta-\tau)(\hat{H}-\mu \hat{N})} a_\mathbf{p} e^{-\tau(\hat{H}-\mu \hat{N})} a_\mathbf{q}^\dagger  \right], \\
Z = \Tr \left[ e^{-\beta(\hat{H}-\mu \hat{N})} \right], 
\end{gather}
where $\mathbf{p,q}$ subscripts label the momentum-dependent spin-orbitals, 
 $a_\mathbf{p}$ and $a_\mathbf{q}^\dagger$ operators are the second-quantized annihilation and creation operators for the corresponding momentum-dependent spin-orbital representation, 
$\tau$ is the imaginary time, 
$\hat{H}$ is the electronic Hamiltonian in the second-quantized representation, 
$\hat{N}$ is the particle-number operator, 
$\mu$ is the chemical potential, 
$\beta$ is the inverse temperature, 
$\Tr$ is the matrix trace taken in the space of all possible configurations in the Fock space, 
$Z$ is the grand-canonical partition function.  

Because of the momentum conservation, the Green's function matrix is block-diagonal in the momentum space and the only non-zero blocks $G_{pq}^\mathbf{k}$ can be labeled with a single momentum index. 
The orbitals used for expressing the Green's function and self-energy matrix elements do not have to be orthogonal, which allows us to use atomic orbitals (AO) directly in the implementation. 
In this work, we use only unrestricted orbital choice (as in unrestricted Hartree--Fock method, UHF), 
where the non-zero (and possibly different) spin blocks are $\alpha\alpha$ and $\beta\beta$, while $\alpha\beta$ and $\beta\alpha$ blocks are zero\cite{note:spin_choice}.  
Originally, the imaginary-time formalism (also known as the Matsubara formalism) was designed for the description of thermal phenomena\cite{Mahan00,Negele:Orland:book:2018,Martin:Interacting_electrons:2016}. 
Although all the analysis performed in this paper is formally temperature-dependent,  
the symmetry-broken Green's function solutions do not change significantly in the broad temperature range, 
effectively corresponding to the zero-temperature limit. The details of this behavior are explained in Ref.\citenum{Pokhilko:local_correlators:2021}. 

The one-particle density matrix $\gamma$ expressed in spin orbitals and any one-electron observable 
$A = \sum_\mathbf{pq} A_\mathbf{pq} a_\mathbf{p}^\dagger a_\mathbf{q}$ can be accessed directly from the imaginary-time Green's function:
\begin{gather}
\gamma_{\mathbf{pq}} = -G_{\mathbf{qp}}(\tau=0^-) \\
\braket{A} = \sum_\mathbf{pq} A_\mathbf{pq} \gamma_\mathbf{pq}. 
\end{gather}
Note that the one-electron operator $A$ can be a particle-number operator with the corresponding integrals $A_\mathbf{pq} = S_\mathbf{pq}$, which are just orbital overlaps.  
Since in solids evaluation of $\braket{A}$ per unit cell is often the objective, a division by the volume of the Brillouin zone is necessary as in Eqs.~\ref{eq:E1b} and \ref{eq:E2b}. 
Because of its simplicity and ansatz-agnostic nature, 
 the analysis of density matrices have been used to understand a broad range of aspects of molecular electronic structure 
through various orbital representations (a far from a complete list of references can be found in Refs\cite{Luzanov:TDM-1:76,Luzanov:TDM-2:79,HeadGordon:att_det:95,Martin:NTO:03,Luzanov:DMRev:12,Dreuw:ESSAImpl:14,Dreuw:ESSAImpl-2:14,Plasser:excitons:2016,Nanda:NTO:17,Wojtek:ImagEx:18,Krylov:Libwfa:18,Dreuw:NTOfeature:2019,Pavel:SOCNTOs:2019,Krylov:Orbitals,Plasser:Visualisation:2019,Nanda:RIXSNTO:20,Wergifosse:respNTO:2020,Nanda:NTO:hyperpolarizability:2021}). 
For the analysis of magnetic effective exchange couplings, various orbital choices have been used\cite{Kahn:book:1993}, 
including natural orbitals\cite{Yamaguchi:magnetic:NO:2000,Malrieu:mag_orbitals:2002,Morokuma:DMRG:biquad_exc:2014,Gagliardi:Cr2muOH:superexchange:2020,Pokhilko:local_correlators:2021,Orms:magnets:17,Kotaru:Fe:SMM:2022}. 
While the most common selection of magnetic orbitals is based on single occupied molecular orbitals, such orbitals are often too delocalized to represent magnetic physics, 
while the natural orbitals remain compact and insightful\cite{Orms:magnets:17}. 

Natural orbitals (NO) for molecules are defined as the eigenvectors of one-particle density matrix;  
their occupancies are the corresponding eigenvalues. 
One may diagonalize a particular spin block of the density matrix and obtain natural spin-orbitals 
or one can diagonalize the spin-averaged density matrix ($\gamma_{\alpha\alpha} + \gamma_{\beta\beta}$) 
and obtain spin-averaged natural orbitals (SA-NO). 
We directly generalize these concepts to solids as
\begin{gather}
\gamma_{\sigma\sigma} \phi_{\mathbf{k}, i, \sigma}^{NO} = d_{\mathbf{k}, i, \sigma} \phi_{\mathbf{k}, i, \sigma}^{NO} \\
(\gamma_{\alpha\alpha}+\gamma_{\beta\beta}) \phi_{\mathbf{k}, i}^{SA-NO} = d_{\mathbf{k}, i} \phi_{\mathbf{k}, i}^{SA-NO}, 
\end{gather}
where $\phi_{\mathbf{k}, i, \sigma}^{NO}$ and $\phi_{\mathbf{k}, i}^{SA-NO}$ are the corresponding natural and spin-averaged natural momentum-dependent orbitals, $d$ denotes their occupation numbers, and the multi-index dependence is expanded explicitly for clarity. 
Because of the momentum dependence, such NOs are extended waves if plotted in the real space. 
To our knowledge, only plane-wave NOs (but not the Bloch NOs as in the eq.\ref{eq:Bloch}) have previously been used in periodic calculations\cite{Kresse:PW_NO:2011,Pucci:NOs_solids:2013}.

Occupancies of natural orbitals can serve to identify open-shell electrons. 
For the cases where a clear separation between doubly occupied, singly occupied, and virtual orbitals is not achieved, 
a number of metrics has been introduced in the past. 
In this work, we use the linear $n_l$ and non-linear $n_{nl}$ indices from the Refs.\cite{Yamaguchi:index:1978,Head-Gordon:Yamaguchi:03}, which are sometimes called Yamaguchi and Head-Gordon indices, respectively
\begin{gather*}
n_l = \sum_p \min(d_p, 2 - d_p), \\
n_{nl} = \sum_p d_p^2 (2-d_p)^2,
\end{gather*}
where $d_p$ is an occupancy of  SA-NO. 
To investigate the open-shell magnetic structure in solids, we generalize these indices for k-dependent SA-NOs:
\begin{gather*}
n_{l,\mathbf{k}} = \sum_p \min(d_{\mathbf{k},p}, 2 - d_{\mathbf{k},p}), \\
n_{nl, \mathbf{k}} = \sum_p d_{\mathbf{k},p}^2 (2-d_{\mathbf{k},p})^2.
\end{gather*}

Imaginary-time-dependent quantities can also be Fourier transformed to a frequency representation. 
Such a set of frequencies is discrete and is called Matsubara frequencies. They are defined respectively for the fermionic and bosonic quantities as: 
\begin{gather}
\omega_n = \frac{(2n+1) \pi}{\beta}, \\
\Omega_n = \frac{2n \pi}{\beta}.  
\end{gather}
In the fully self-consistent GW, during each of the iterations, we solve the Dyson equation:
\begin{gather}
G^{-1}(i\omega_n) = G^{-1}_0(i\omega_n) - \Sigma[G](i\omega_n),
\protect\label{eq:Dyson}
\end{gather}
where $G$ is the full one-particle Green's function, 
$G_0$ is the Green's function of independent electrons, 
and $\Sigma[G]$ is the self-energy that has a functional dependence only on the full Green's function $G$ and two-electron integrals. 
The self-energy encodes all the orbital relaxation, screening, and correlation effects. 
In particular, fully self-consistent approximations are based on perturbative expansions of 
the self-energy, preserving the functional dependence on $G$ and the two-electron integrals. 
The simplest example of the such an approximation is the Hartree--Fock theory. 
In the notation above, the Hartree--Fock self-energy is only static (does not depend on time) 
and is a part the Fock matrix times minus one. 
Correlated post-Hartree--Fock methods have an additional dynamical part of the self-energy,
describing correlation and screening.

In this work, we use the fully self-consistent GW approximation\cite{Hedin65,G0W0_Pickett84,G0W0_Hybertsen86,GW_Aryasetiawan98,Stan06,Koval14,scGW_Andrey09,Kutepov17,Iskakov20,Yeh:GPU:GW:2022}. 
The theoretical and implementation details for solids are fully covered in Refs.\cite{Iskakov20,Yeh:GPU:GW:2022}. 

The imaginary time dependence of the Green's function and self-energy are very important since  
they provide a route for evaluating properties and electronic energy. 
For example, electronic energy is evaluated from the Galitskii--Migdal expression as
\begin{gather}
E_{1b} = \frac{1}{V_{BZ}}\sum_{pq} h_{pq}^\mathbf{k} \gamma_{pq}^\mathbf{k},  
\protect\label{eq:E1b} \\
E_{2b} = \frac{1}{2V_{BZ} \beta}\sum_{\omega_n} \sum_{pq} G_{qp}^\mathbf{k}(\omega_n) \Sigma_{pq}^\mathbf{k}(\omega_n), 
\protect\label{eq:E2b}
\end{gather}
where $h_\mathbf{pq}$ are one-electron integrals, $V_{BZ}$ is the volume of a Brillouin zone, 
$E_{1b}$ and $E_{2b}$ are the one- and two-body parts of the total energy. 

Analysis based on NOs has some limitations. 
It can give insights only into the structure of the one-particle density matrix, 
but not into the two-particle observables. 
This limitation explains why NOs are not sufficient for a complete understanding of energy differences 
and, therefore, effective exchange couplings. 
To mitigate this issue, we also include two-particle observables in our analysis. 
Recently, we applied the thermodynamic Hellmann--Feynman theorem and showed that for any fully self-consistent Green's function method the two-particle density matrices have disconnected and connected parts\cite{Pokhilko:tpdm:2021}: 
\begin{gather}
\Gamma_{\braket{\mathbf{p q} | \mathbf{r s}}} = 
\Gamma^\text{disc}_{\braket{\mathbf{p q} | \mathbf{r s}}} + \Gamma^\text{conn}_{\braket{\mathbf{p q} | \mathbf{r s}}}, \\
\Gamma^\text{disc}_{\braket{\mathbf{p q} | \mathbf{r s}}} = \gamma_{\mathbf{pr}} \gamma_{\mathbf{qs}} - \gamma_{\mathbf{ps}} \gamma_{\mathbf{qr}}.  
\protect\label{eq:Gamma}
\end{gather}
Such two-particle density matrices fully reproduce two-particle part of the energy when a trace with integrals is taken:
\begin{gather}
E_{2b} = \frac{1}{V_{BZ}^2}\sum_\mathbf{pqrs} \braket{\mathbf{pq} | \mathbf{rs}} \Gamma_{\braket{\mathbf{p q} | \mathbf{r s}}}.
\end{gather}
The explicit expression for the connected part of the two-particle density matrix (also called an electronic cumulant) 
depends on a particular approximation used, 
but it is always a contraction between a 4-point vertex function (encoding effective interactions) 
and Green's functions at all 4 points.  
The fully self-consistent GW cumulant\cite{Pokhilko:tpdm:2021} is defined as
\begin{gather}\label{eq:gw_cummulant}
\Gamma_{(\mathbf{p_0 q_0} | \mathbf{r_0 s_0})}^\text{GW \ conn} = 
-\frac{1}{\beta}\sum_{\Omega_m} 
\sum_{\mathbf{pqrs}} \Pi_{\mathbf{r_0 s_0 pq}}(\Omega_m) W_{(\mathbf{pq}|\mathbf{rs})}(\Omega_m) \Pi_{\mathbf{rs p_0 q_0}}(\Omega_m), 
\end{gather}
where $\Pi$ is a polarization function. 
The physical meaning of this expression becomes transparent if one generalizes it to the full two-particle Green's function, describing propagation of two particles. When particles propagate in a medium, they interact with it; therefore, their propagators become renormalized. 
Propagation of two particles in the medium can happen either without interaction with each other (giving the disconnected terms) or with interaction with each other. The structure of such an interaction is described by the 4-point vertex function, which is just $W$ for GW. 

Due to the equivalence and possibility of evaluating the electronic energy using either frequency-dependent formalism (Galitskii-Migdal) or two-particle density matrix traced with the integrals,
it is sufficient to investigate only one route. 
Consistently, with our previous work for molecules\cite{Pokhilko:local_correlators:2021}, 
we choose to investigate only local two-particle observables for solids. 
Kubo\cite{Kubo:cumulant:1962} introduced statistical cumulants of two operators $\hat{X}$ and $\hat{Y}$, measuring their covariance:
\begin{gather}
\braket{\delta\hat{X} \delta\hat{Y}} =
 \braket{\hat{X}\hat{Y}} - \braket{\hat{X}}\braket{\hat{Y}} = 
\braket{(\hat{X} - \braket{\hat{X}}) (\hat{Y} - \braket{\hat{Y}})}.
\end{gather}
In localized orbitals, the two-particle charge\cite{Ruedenberg:chem_bond:1962,Jorge:bond_index:1985,Torre:popul:cumulants:2002,Bochicchio:bond_order:2003,Goddard:corr_chem_bond:1998,Luzanov:bond_indices:2005,Mayer:bond_order:2007} 
and spin \cite{Davidson:local_spin:2001,Davidson:local_spin:2002,Davidson:MolMagnets:2002,Hess:local_spin:2005,Luzanov:SpinCorr:15,Pokhilko:BS-GW:solids:2022} cumulants have been extensively used to quantify bond orders, charge-transfer, and covalent nature of electronic states, as well as magnetic properties. 
In this work, we extend the concept of a charge cumulant to solids. 
We define a local particle-number operator as 
\begin{gather}
\hat{N}_A^\mathbf{k} = \sum_{pq \in A} S_{pq}^\mathbf{k} ( a^\dagger_{\mathbf{k}, p,\alpha} a_{\mathbf{k},q,\alpha} + a^\dagger_{\mathbf{k},p,\beta} a_{\mathbf{k},q,\beta}). 
\end{gather}
Similarly to spin correlators\cite{Pokhilko:BS-GW:solids:2022} $SS_{AB}^{\mathbf{k}=0}$ and $SS_{AB}$, we introduce particle-number correlators $NN_{AB}^{\mathbf{k}=0}$ and $NN_{AB}$:
\begin{gather}
NN_{AB}^{\mathbf{k}=0} = \braket{\delta \hat{N}_A^{\mathbf{k}=0} \delta \hat{N}_B^{\mathbf{k}=0}},  \\
NN_{AB} = \frac{1}{V_{BZ}}\sum_\mathbf{k} \braket{\delta \hat{N}_A^\mathbf{k} \delta \hat{N}_B^\mathbf{k}}. 
\end{gather}
Such correlators quantify charge fluctuations and covariances between localized subsets of orbitals $A$ and $B$. 
$NN_{AB}^{\mathbf{k}=0}$ measures charge covariances at a $\Gamma$-point ($\mathbf{k} = 0$); 
$NN_{AB}$ measures charge covariances in the real-space representation between orbital subsets in the same unit cell. 
Both measures provide diagnostics of the charge-transfer magnitude for a given Green's function solution. 
Due to high computational cost of evaluating the connected component of the two-particle density matrix, we include it only for very few $\mathbf{k}$-points. 
In this work, we neglect the connected contribution from the cumulant and consider only the disconnected part. 
Such an approximation is permissible since GW is a weakly correlated method and the magnitude of the electronic cumulant is predicted to be rather small. 
Previously, we tested the validity of this assumption for spin correlators\cite{Pokhilko:BS-GW:solids:2022}.

\section{Results and discussion}
\subsection{Computational details}
\begin{figure}[!h]
  \includegraphics[width=8cm]{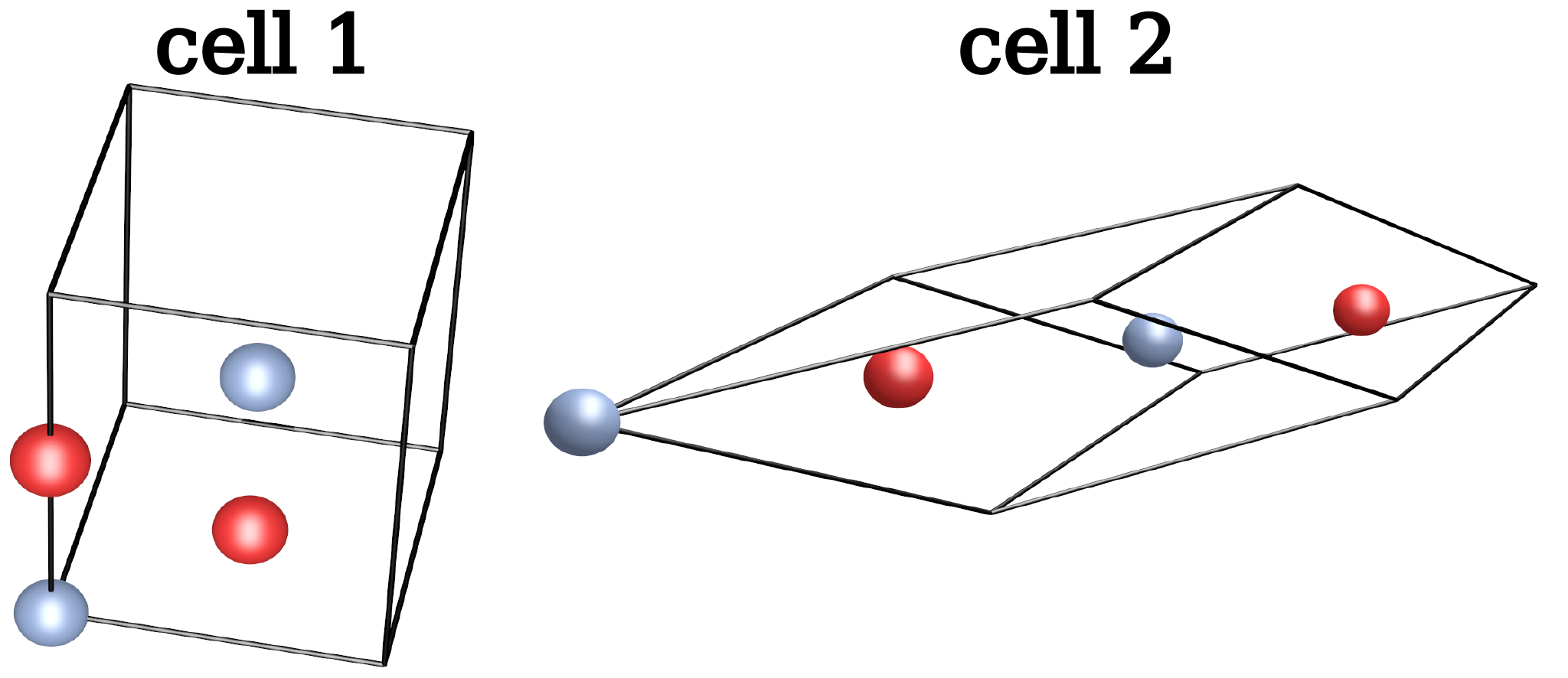}
\centering
\caption{Two supercells used in this paper to capture solutions of different types.  
         \protect\label{fig:lattice}
}
\end{figure}
We performed calculations in the same manner as before in Ref.~\cite{Pokhilko:BS-GW:solids:2022,Iskakov20}. 
We used \emph{gth-dzvp-molopt-sr} basis set\cite{GTHBasis}, \emph{gth-pbe} pseudopotential\cite{GTHPseudo}, and
\emph{def2-svp-ri} auxiliary basis\cite{RI_auxbasis} for
the resolution-of-identity approximation (RI) of two-electron integrals. 
We used the rock-salt NiO, CoO, FeO structures with 
the lattice constants a = 4.1705\AA~\cite{Morosin:NiO:exchange_striction:1971}, 
4.2630\AA~\cite{Takeuchi:CoO:1979}, 4.285\AA~\cite{Crisan:FeO:2011}, respectively.
We used supercells of different types to capture solutions of different nature (Fig.~\ref{fig:lattice}). 
In each of the cells, we obtained ferromagnetic and antiferromagnetic broken-symmetry Green's function solutions using frequency-dependent CDIIS algorithm for convergence acceleration from Ref.\cite{Pokhilko:algs:2022}. The GW calculations were started 
from the corresponding zero-temperature UHF solutions.  
We denote the ferromagnetic solutions with the highest spin projection as HS 
and the antiferromagnetic broken-symmetry solutions as BS.  
The Monkhorst--Pack k-point grid was used for the Brillouin-zone sampling\cite{Monkhorst:Pack:k-grid:1976}. 
The finite-size effects were accounted for within 
the exchange terms in UHF and GW with the probe-charge-corrected Ewald approach\cite{EwaldProbeCharge,CoulombSingular}. 
All the frequency-dependent quantities were written in the intermediate representation\cite{Yoshimi:IR:2017}
with $\Lambda = 10^5$ and 136 functions. 
The one- and two-electron integrals were produced by the PySCF code\cite{PYSCF}. 
RI was used for all the calculations performed with the local in-house Green's function code \cite{Rusakov16,Iskakov20,Pokhilko:tpdm:2021,Pokhilko:local_correlators:2021,Yeh:X2C:GW:2022,Yeh:GPU:GW:2022,Pokhilko:algs:2022}. 
Since we use very compact basis sets in solids, we used Mulliken-like partitioning of orbitals with simple AOs. 
Such a partitioning choice allows us to avoid costly integral transformations for solids. 
Natural orbitals are visualized with Gabedit 2.5.1\cite{Gabedit:2011} as isosurfaces with isovalues $0.050$ 
and rendered with POV-Ray 3.7.0\cite{povray}.  
The $k$-path was constructed with ASE 3.22.0 package\cite{ase-paper} and interpolated with Wannier interpolation in AOs. 
The largest eigenvalues of the interpolated density matrix slightly exceeded 2 at some k-points 
due to an imperfect interpolation;  
in such cases we capped these occupancies by 2 in the evaluation of $n_l$ and $n_{nl}$ indices. 
We use the symmetry convention for special points and for the $k$-path according to Ref. \cite{Curtarolo:kpath:symmetry:2010}.

\subsection{Goodenough--Kanamori rules}
In the compounds considered, the effective magnetic Hamiltonian\cite{Martin:NiO:exchange:2002,Majumdar:NiO:MnO:DFT:J:2011,Pokhilko:BS-GW:solids:2022,Pokhilko:Neel_T:2022} has only non-zero couplings between the nearest neighbors $J_1$ and between the next-nearest neighbors $J_2$ and is expressed as
\begin{gather}
H = -J_{1} \sum_{\braket{i,j}} \vec{S}_i \vec{S}_j - J_{2} \sum_{\braket{\braket{i,j}}}  \vec{S}_i \vec{S}_j. 
\protect\label{eq:Ham_defs}
\end{gather}
Semiempirical Goodenough--Kanamori (GK) rules\cite{Goodenough:direct_exchange:1960,Kanamori:exchange_mechanisms:1959,GK:rules:summary} are commonly used to explain the magnetic origin of effective interactions in transition-metal oxides and related compounds. 
They are based on simplistic models and assume localized interactions between orbitals and their overlaps. 
In particular, they consider the $d-p-d$ bonds and 
show that the 90$^\circ$ angle between Ni--O--Ni results in a ferromagnetic interaction, 
while the 180$^\circ$ angle results in the antiferromagnetic interaction. 
With our tools for the analysis of electronic structure, we can check the validity of GK assumptions directly from electronic structure calculations without any additional approximations since natural orbitals can directly explain which orbitals are on average occupied by the open-shell electrons. Multielectron multicenter bonding from Ref.\cite{Halpern:superexchange:1966} 
provide an alternative description of magnetic interactions in terms of delocalized orbitals; 
natural orbitals directly generalize these ideas.   
Because of the local assumption, the GK rules predict that the open-shell electrons occupy orbitals with the same composition of atomic orbitals. 
As a result, all the corresponding NOs would be composed of $d-p-d$ orbitals times the k-dependent phases with the same combinations of AOs from the unit cell. 
In particular, such local contributions from nickel atoms in NiO are attributed to local $t_{2g}$ atomic orbitals 
and contributions from oxygen are attributed exclusively to $p$-orbitals.  
\begin{figure}[!h]
  \includegraphics[width=4cm]{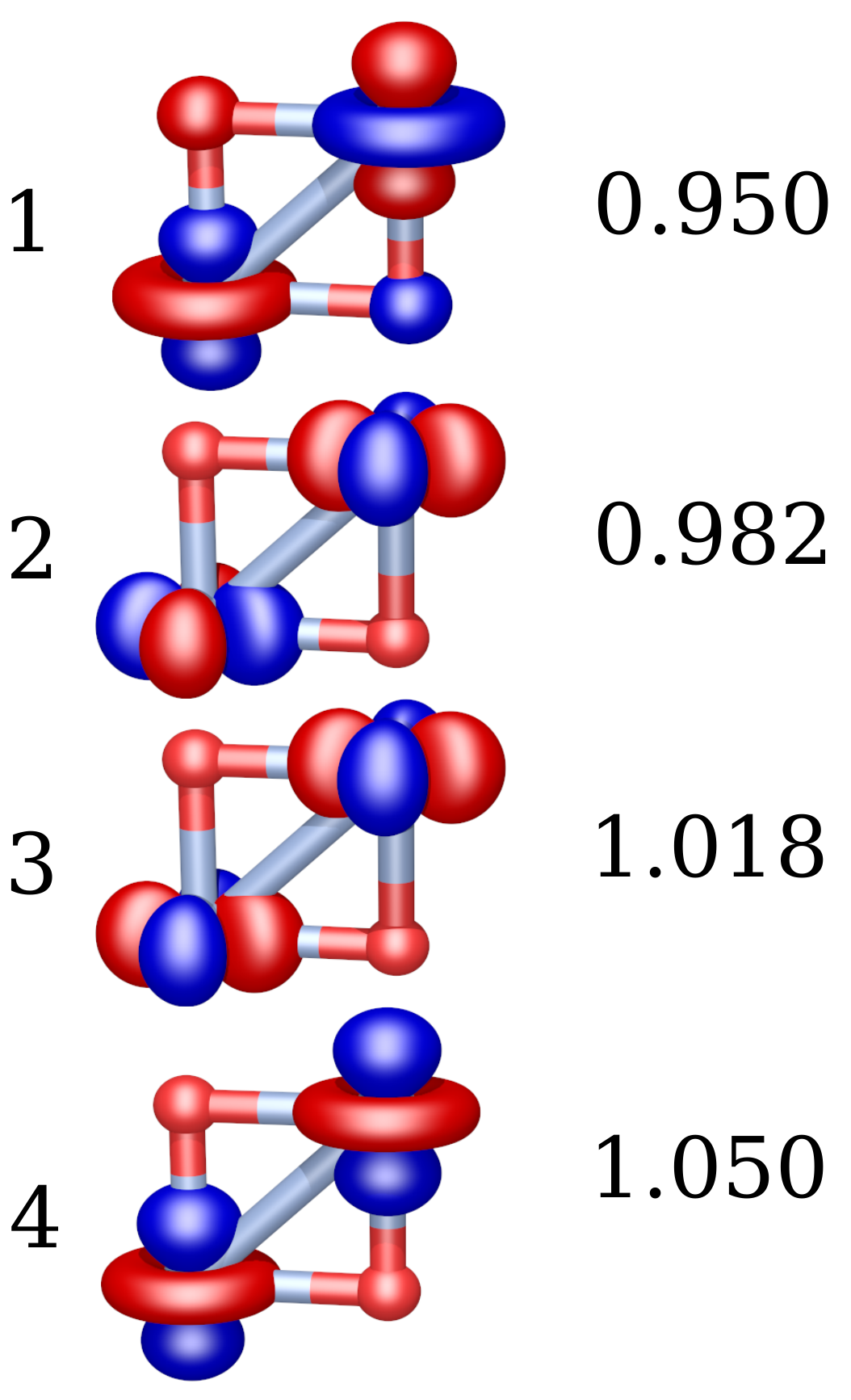}
  \includegraphics[width=4cm]{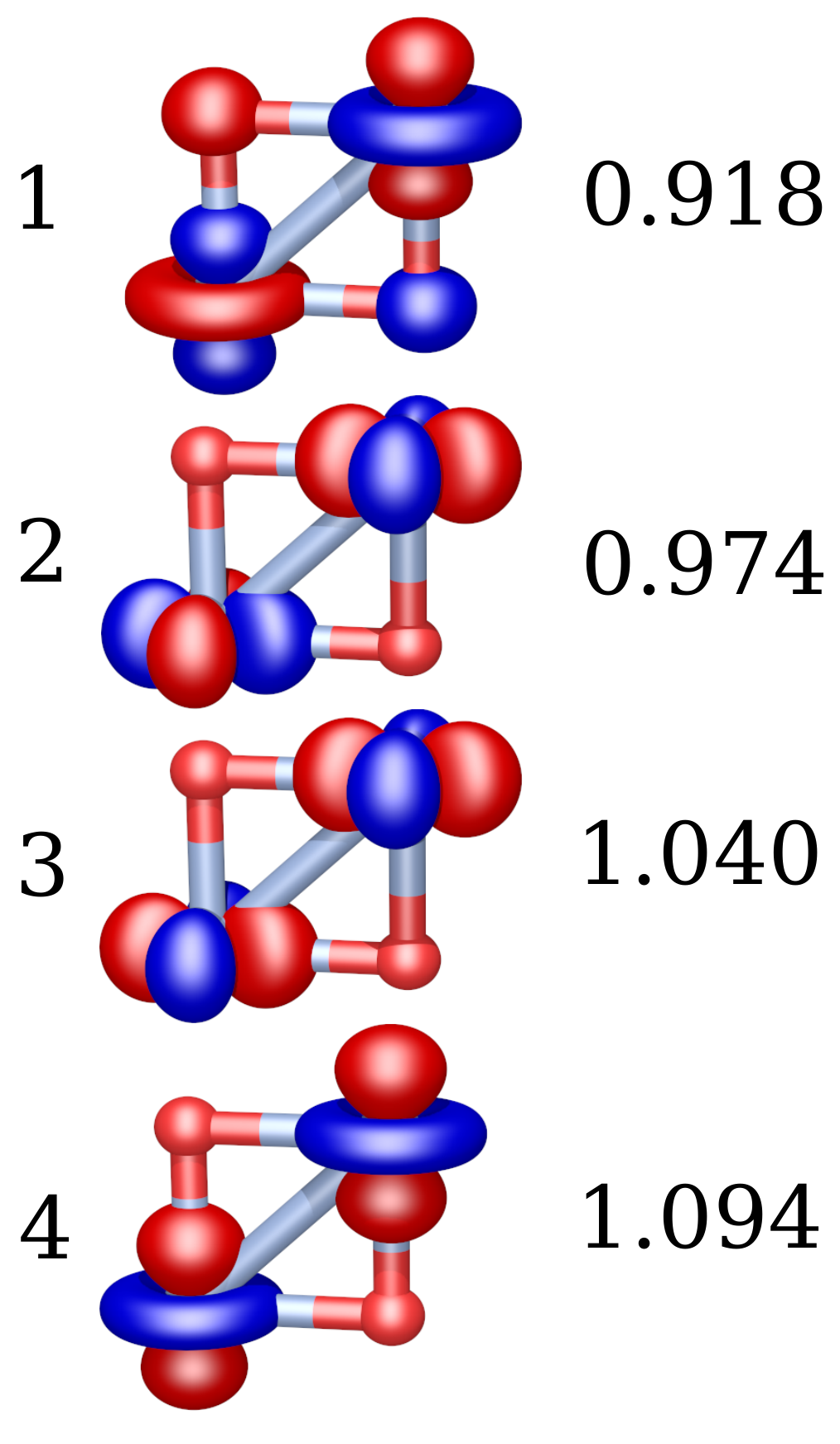}
\centering
\caption{SA-NOs and their occupancies computed for the broken-symmetry solutions at the $\Gamma$ point 
         of the NiO \textbf{cell 1} within UHF (left) and GW (right).  
         The following atoms are pictured in the unit cell, 
         starting from left top corner and moving clockwise, O-Ni-O-Ni. 
         For every orbital, its number is shown on the left and its occupancy is shown on the right. 
         \protect\label{fig:NiO_90_orbs} 
}
\end{figure}
\begin{figure}[!h]
  \includegraphics[width=7cm]{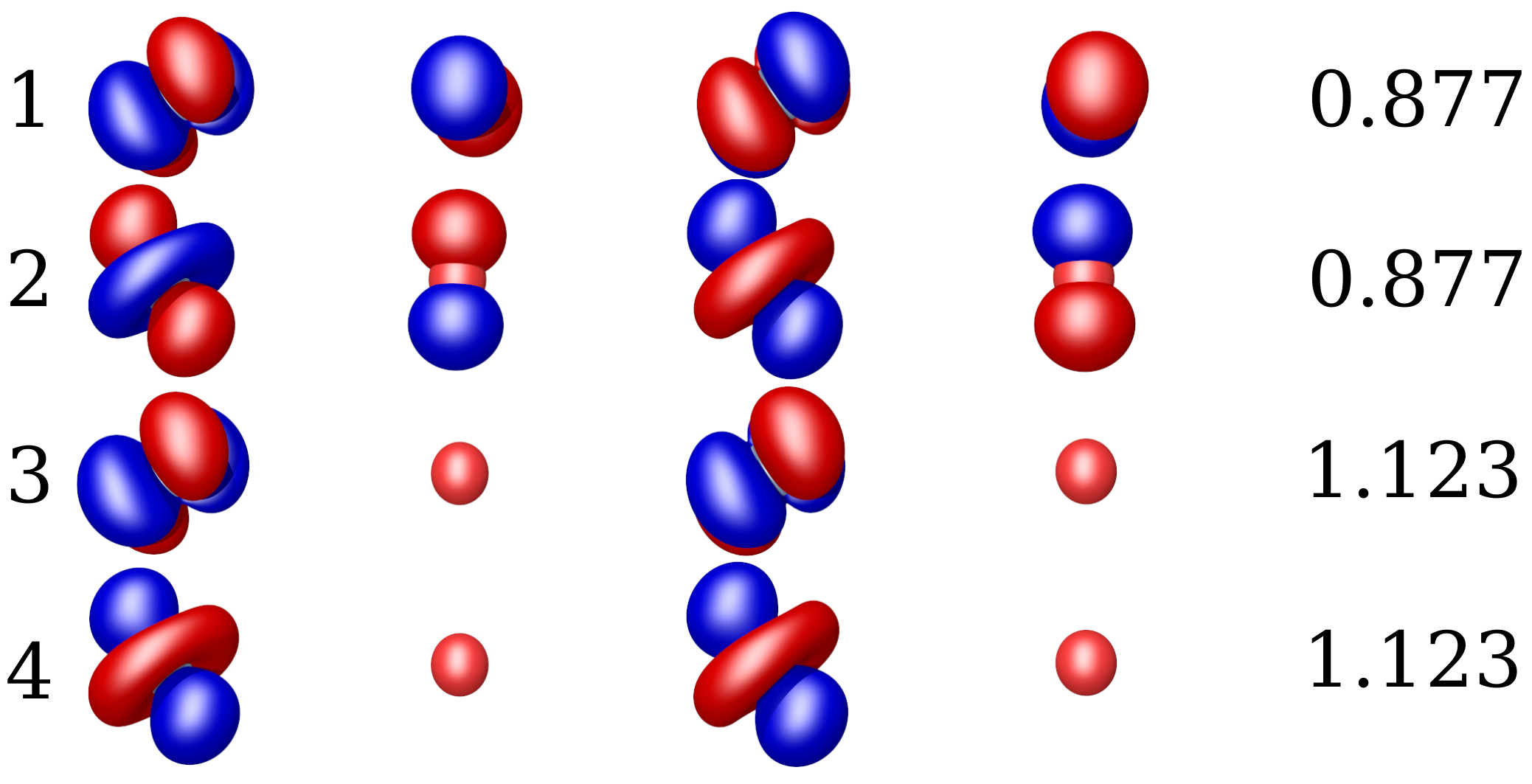}
\hfill
  \includegraphics[width=7cm]{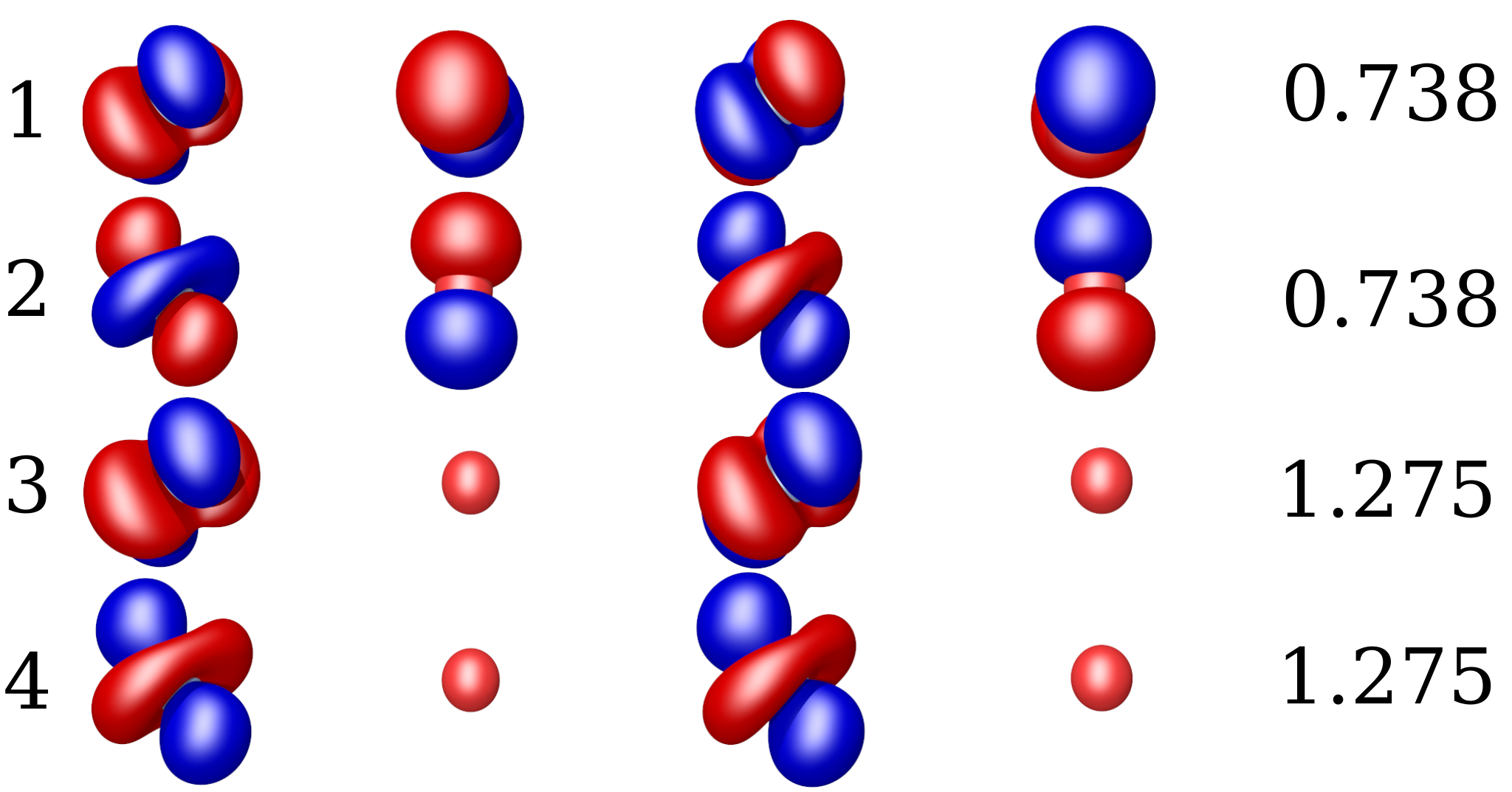}
\centering
\caption{SA-NOs and their occupancies computed for the broken-symmetry solutions at the $\Gamma$ point 
         of the NiO \textbf{cell 2} within UHF (left) and GW (right). 
         To see a partly antibonding character of the d-p-d combinations, 
         one can apply translations with respect to $\mathbf{k}=0$ wave, 
         shown in Fig.~S1 in SI.
         The following atoms are pictured in the unit cell, starting from left to right, Ni-O-Ni-Ni. 
         For every orbital, its number is shown on the left and its occupancy is shown on the right. 
         \protect\label{fig:NiO_60_orbs}
}
\end{figure}

\subsection{Natural orbitals at $\Gamma$ point}
However, the computed natural orbitals show a different picture. 
The frontier spin-averaged natural orbitals with occupancies close to 1 are separated well from the 
doubly occupied and virtual SA-NOs. 
The frontier SA-NOs of NiO computed with UHF and GW for the broken-symmetry solutions at a $\Gamma$ point are shown in Figs~\ref{fig:NiO_90_orbs} and \ref{fig:NiO_60_orbs} for \textbf{cell 1} and \textbf{cell 2}. 
Little groups at the $\Gamma$ point of the considered structures 
have a high symmetry preserving non-Abelian structure of rotations\cite{Chen:rep:space_groups:1985} 
and leading to degenerate irreducible representations, 
manifestations of which we observe as degenerate occupancies.  
Qualitatively, the frontier SA-NOs are very different from the ones expected from GK rules. 
First, SA-NO \# 1 in \textbf{cell 1} (Fig~\ref{fig:NiO_90_orbs}) has large 
contributions from $s$-orbitals on oxygen, but the contribution from the $p$-orbitals is zero.  
While the Kanamori's paper \cite{Kanamori:exchange_mechanisms:1959} has a remark that $s$-orbitals do not violate the symmetry arguments leading to ferromagnetic interactions between nearest neighbors 
if the overlap between $s$- and $d$-orbitals is zero, 
they are not considered in the follow-up publications. 
The $s$- and $d-$orbitals composing SA-NO \# 1 have a non-zero overlap, which makes the analysis non-trivial. 
The $d$-orbitals from the SA-NOs form symmetric and antisymmetric combinations. 
The net sum of occupations of the corresponding symmetric and antisymmetric SA-NOs occupations is close to 2, 
meaning that the electrons partly transfer from the antisymmetric SA-NOs (occupations are slightly smaller than 1) to the symmetric ones (occupations are slightly larger than 1), leading to magnetic interactions, 
since due to the Pauli repulsion such a redistribution of electrons with the same spin in the ferromagnetic solutions 
cannot happen.  

SA-NOs \# 1 and \# 2 in \textbf{cell 2} (Fig. \ref{fig:NiO_60_orbs}) have contributions from $p$-orbitals, 
but their orientation does not coincide with the $d$-orbitals, which becomes especially clear if the translation to the nearest-neighbors is applied (Fig.~S1 in SI). 
This misalignment, resulting only  in partial antibonding interactions, is also a deviation from the GK rules. 
In \textbf{cell 2}, SA-NOs \# 1 and \# 2 (partly antibonding orbitals) are partly depopulated;  
SA-NOs \# 3 and \# 4 (non-bonding orbitals) are partly populated, while the net sum of occupations between the corresponding symmetric and antisymmetric combinations is again close to 2.  
The occupancies of the corresponding bonding/antibonding SA-NOs explain the origin of partial charge transfer from the oxygen to the nickel atom.

\subsection{Natural orbitals at other k-points}
\begin{figure}[!h]
  \includegraphics[width=15cm]{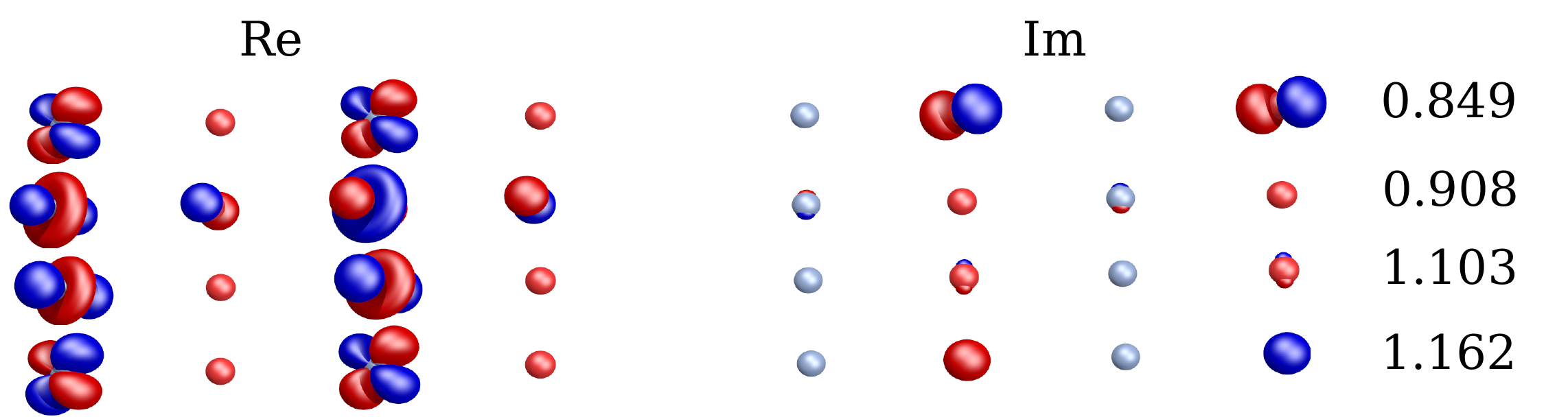}
\centering
\caption{SA-NOs and their occupancies computed from the broken-symmetry solutions at $\vec{k}=(0.3986,0.080,-0.8770)$ 
         of NiO in \textbf{cell 2} within GW ($\vec{k}$ is written in the absolute notation). 
         Both real (left) and imaginary (right) parts of the SA-NOs are visualized. 
         \protect\label{fig:NiO_60_orbs_7k}
}
\end{figure}
Characters and occupations of SA-NOs depend strongly on momentum, 
which is again a violation of the prediction of the GK rules.  
Fig. \ref{fig:NiO_60_orbs_7k} shows the frontier SA-NOs of NiO at one selected momentum point, 
where $s$-orbitals contribute to the imaginary part of the SA-NO. 
Due to a mixed character, 
interpretation of such orbitals in terms of bonding and antibonding orbitals is less straightforward.  
Nonetheless, even in this case the net sum of occupancies for the corresponding mixed bonding/antibonding pairs of SA-NOs is close to 2. 
This remains true for other k-points.  
\begin{figure}[!h]
  \includegraphics[width=7cm]{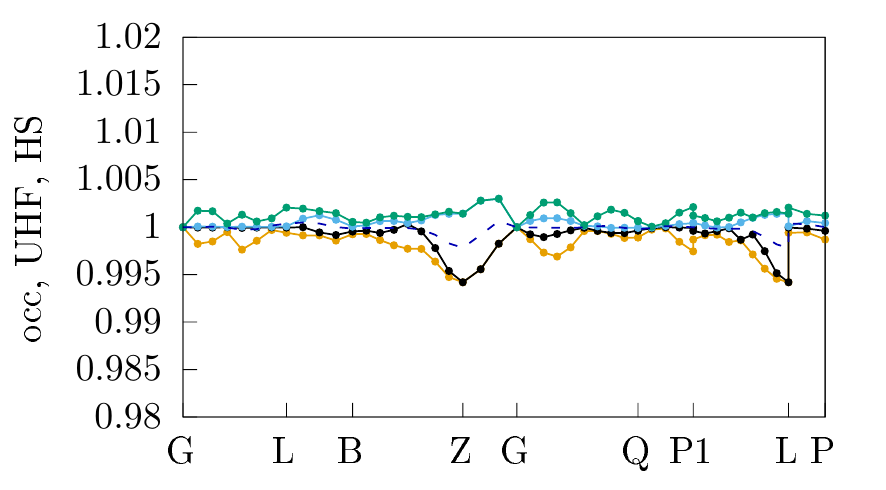}
\hfill
  \includegraphics[width=7cm]{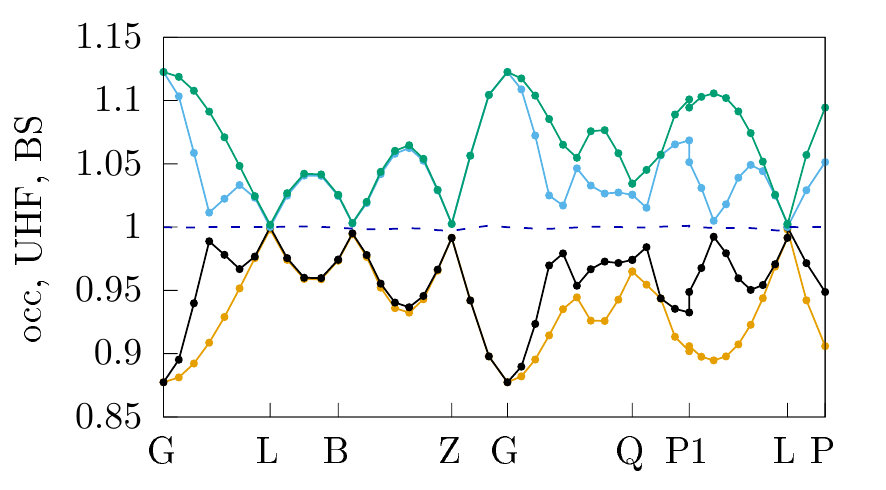} \\
  \includegraphics[width=7cm]{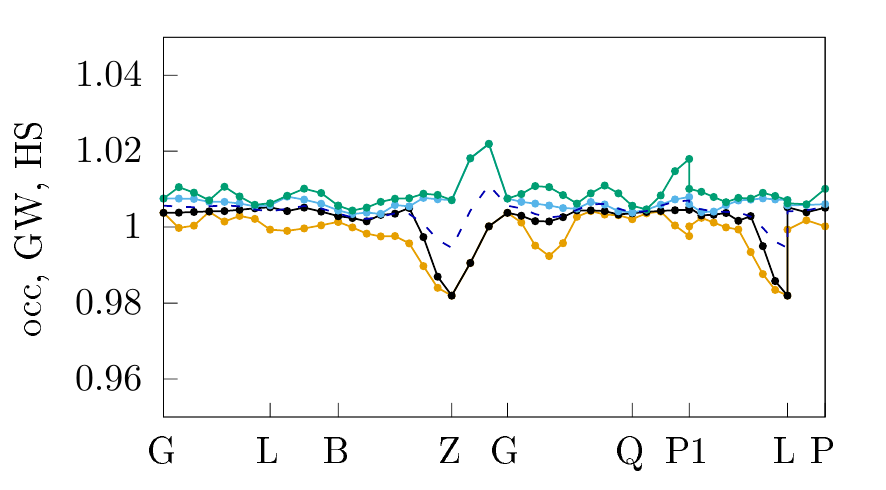}
\hfill
  \includegraphics[width=7cm]{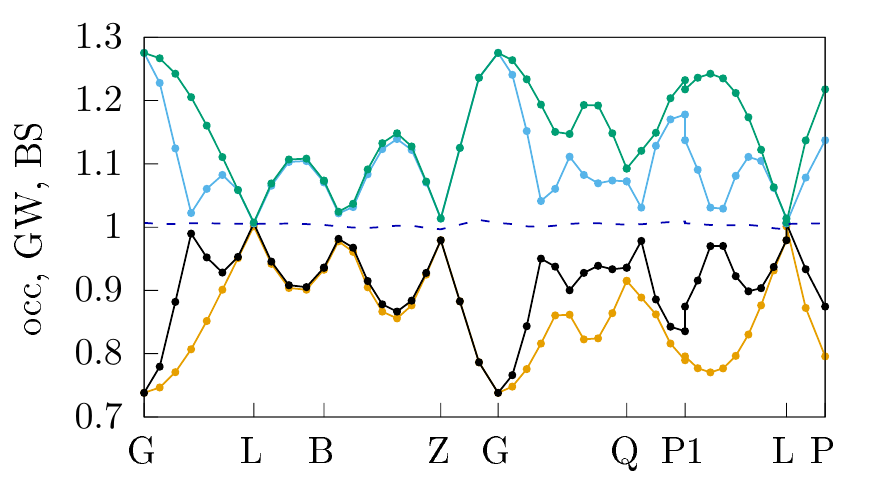} 
\centering
\caption{Occupation numbers of the frontier SA-NOs for the k-points along the interpolated path for 
         NiO in \textbf{cell 2} evaluated with UHF (top) and GW (down) with $5\times 5\times 5$ grid for HS (left) 
         and BS (right) solutions. 
         The dashed blue line is the average of the occupancies of all 4 frontier SA-NOs at each k-point. 
         \protect\label{fig:NiO_60_occ}
}
\end{figure}
\begin{figure}[!h]
  \includegraphics[width=7cm]{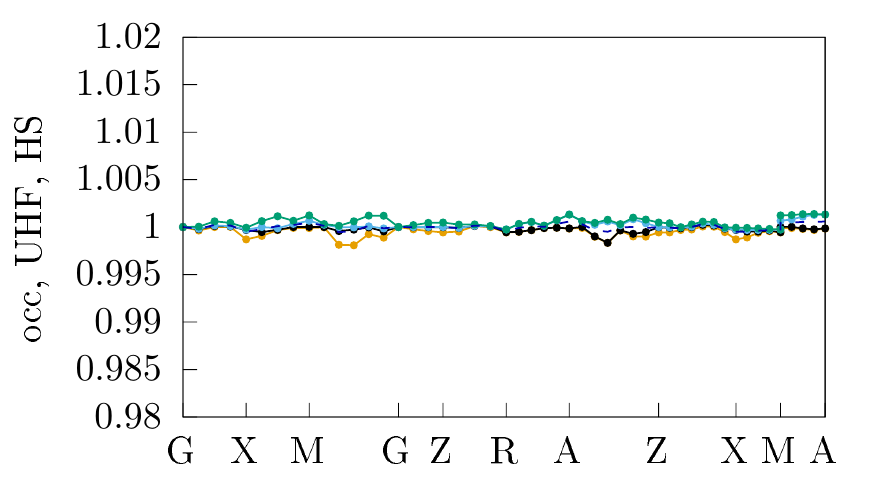}
\hfill
  \includegraphics[width=7cm]{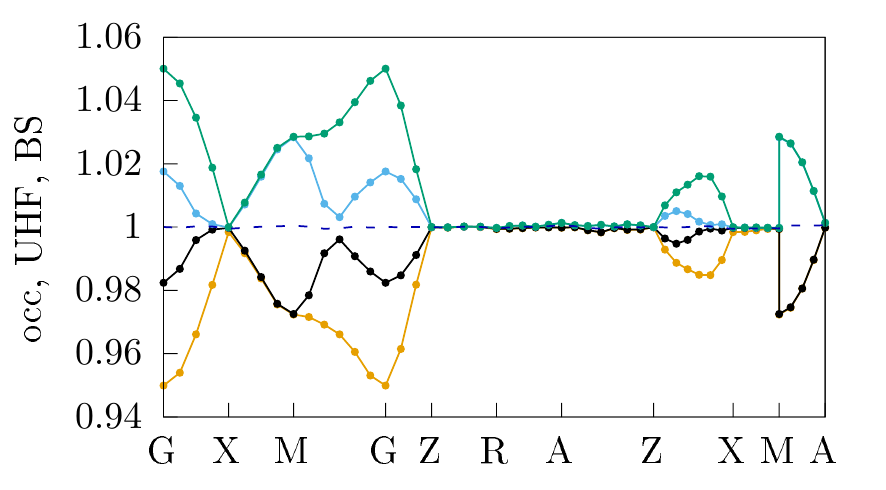} \\
  \includegraphics[width=7cm]{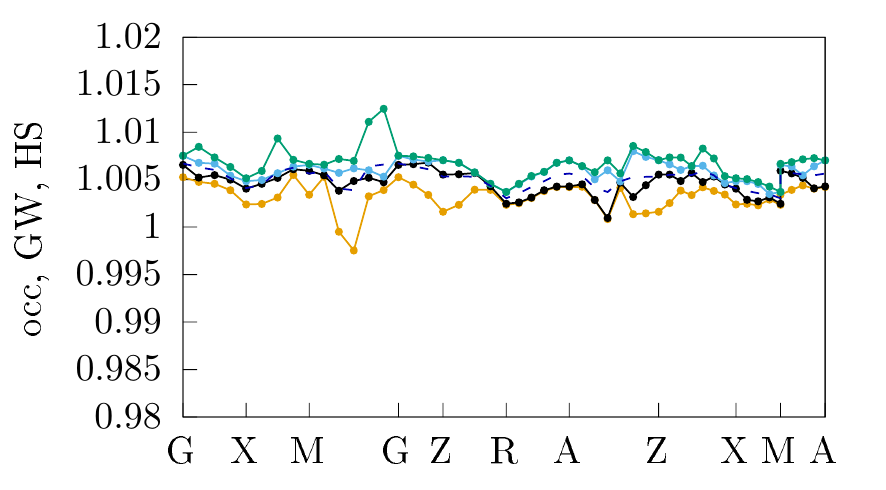}
\hfill
  \includegraphics[width=7cm]{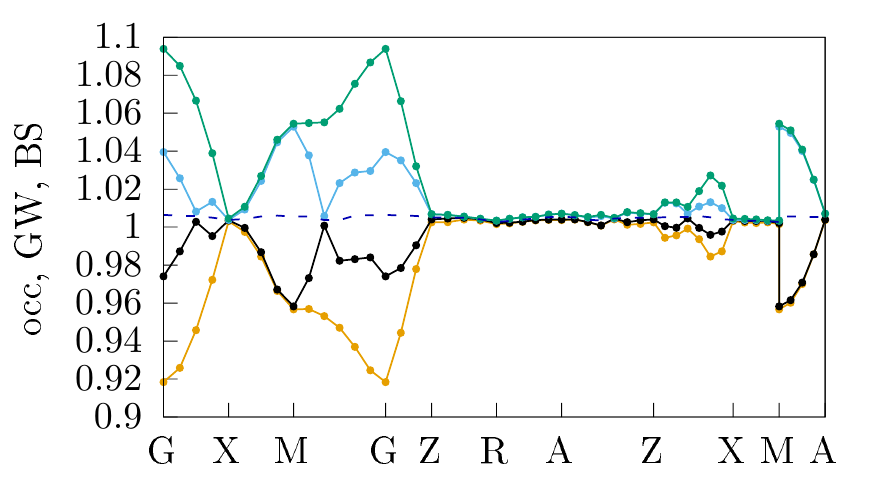} 
\centering
\caption{Occupation numbers of the frontier SA-NOs for the k-points along the interpolated path for 
         NiO in \textbf{cell 1} evaluated with UHF (top) and GW (down) with $5\times 5\times 5$ grid for HS (left) 
         and BS (right) solutions. 
         The dashed blue line is the average of the occupancies of all 4 frontier SA-NOs at each k-point. 
         \protect\label{fig:NiO_90_occ}
}
\end{figure}
Fig.~\ref{fig:NiO_60_occ} shows occupancies of SA-NOs along the Wannier interpolation path (k-path) in rhombohedral (type 1) \textbf{cell 2}. 
The SA-NO occupancies of the HS solution along the k-path  are very flat and close to 1,  
meaning that the structure of the ferromagnetic solution is similar across different k-points 
(but the underlying composition of SA-NOs, of course, depends on momentum). 
The SA-NO occupancies of the BS solution show deviations from 1. 
Interestingly, these occupation deviations are almost perfectly symmetric with respect to 
the average occupation of these SA-NOs, which is almost a constant close to 1 along the entire k-path. 
This behavior is preserved when electron correlation is included by means of GW, 
implying that the main contributions to the differences between the solutions come from only 
these 4 frontier SA-NOs at each k-point. These general observations are also true for the occupancies of frontier SA-NOs at \textbf{cell 1} (Fig.~\ref{fig:NiO_90_occ}), but the deviations from 1 are much smaller. 
All these patterns of occupation numbers along the corresponding $k$-paths are also preserved for CoO and FeO with 6 and 8 open-shell frontier SA-NOs, respectively, shown in Figs. S4, S5, S10, and S11 in SI. 
\begin{figure}[!h]
  \includegraphics[width=8cm]{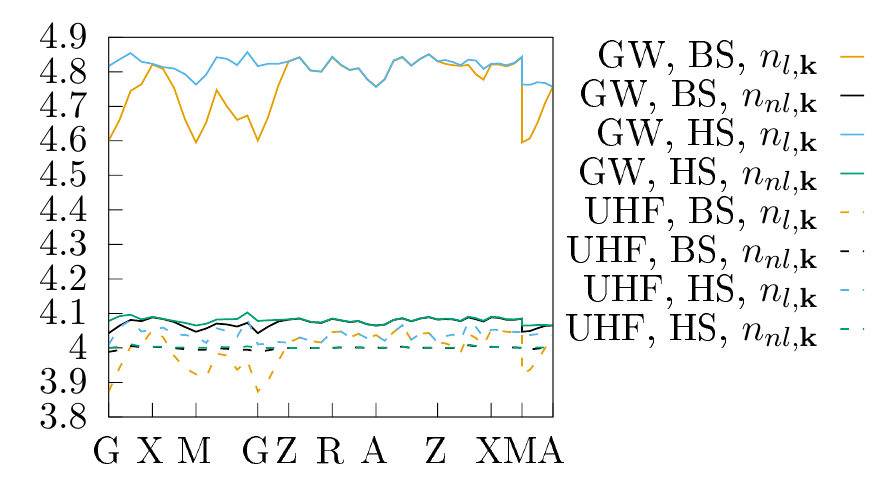}
\hfill
  \includegraphics[width=8cm]{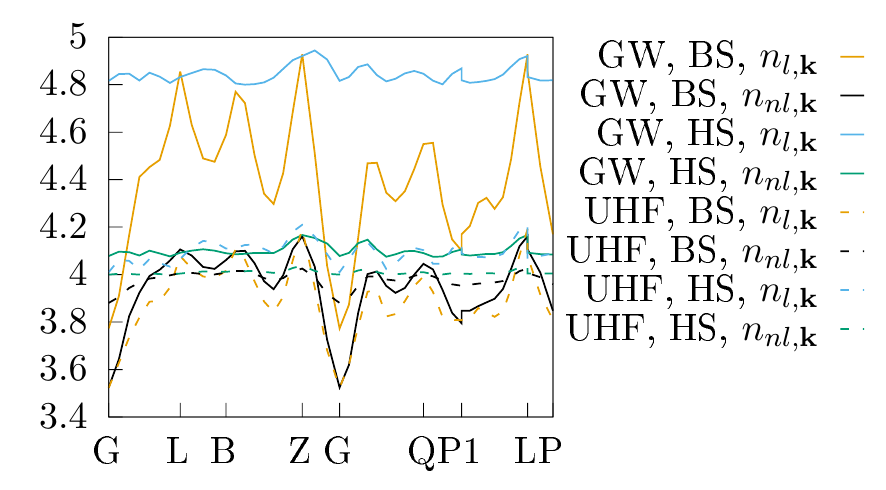} 
\centering
\caption{Effective numbers of open-shell electrons for the UHF (dashed lines) and GW (solid lines) solutions in \textbf{cell 1} (left) and \textbf{cell 2} (right). 
         \protect\label{fig:NiO_indices}
}
\end{figure}

Fig.~\ref{fig:NiO_indices} shows the effective number of open-shell electrons of NiO along the interpolation 
k-path in both cells. 
For UHF, $n_{l,\mathbf{k}}$ and $n_{nl,\mathbf{k}}$ agree well and give an estimate close to 4. 
The nonlinear index, $n_{nl,\mathbf{k}}$, shows smaller deviations from 4 along the k-path. 
The difference between the indices increases when GW dynamic correlation is included 
since $n_{l,\mathbf{k}}$ is more susceptible to a large number of small occupancies due to the presence of dynamic correlation. 
These observations are fully consistent with $n_{l}$ and $n_{nl}$ behavior for molecules\cite{Head-Gordon:Yamaguchi:03}. 
In \textbf{cell 2}, $n_{nl,\mathbf{k}}$ drops at a $\Gamma$ point to a minimum along the k-path; there are also additional local minima. All these minima correspond to extremal differences in occupancies in Fig.~\ref{fig:NiO_60_occ}. 
In \textbf{cell 1}, $n_{nl,\mathbf{k}}$ is almost flat, while $n_{l,\mathbf{k}}$ is sensitive enough to recognize smaller differences in occupancies in Fig.~\ref{fig:NiO_90_occ}.  
Nonetheless, we notice that simple occupancies of the frontier SA-NOs are much easier to interpret than rather sophisticated indices $n_{l,\mathbf{k}}$ and $n_{nl,\mathbf{k}}$. 
\begin{figure}[!h]
  \includegraphics[width=8cm]{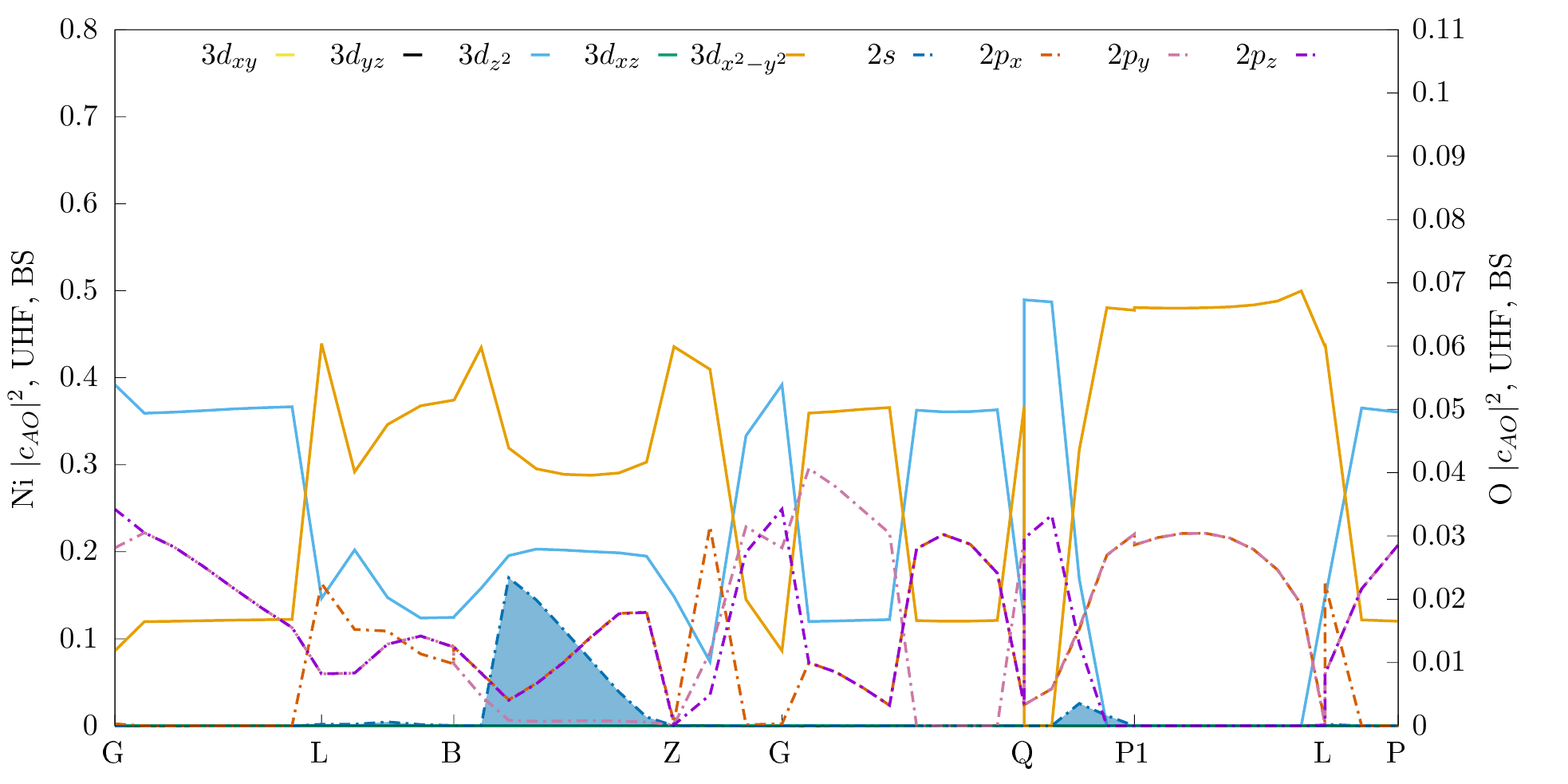}
\hfill
  \includegraphics[width=8cm]{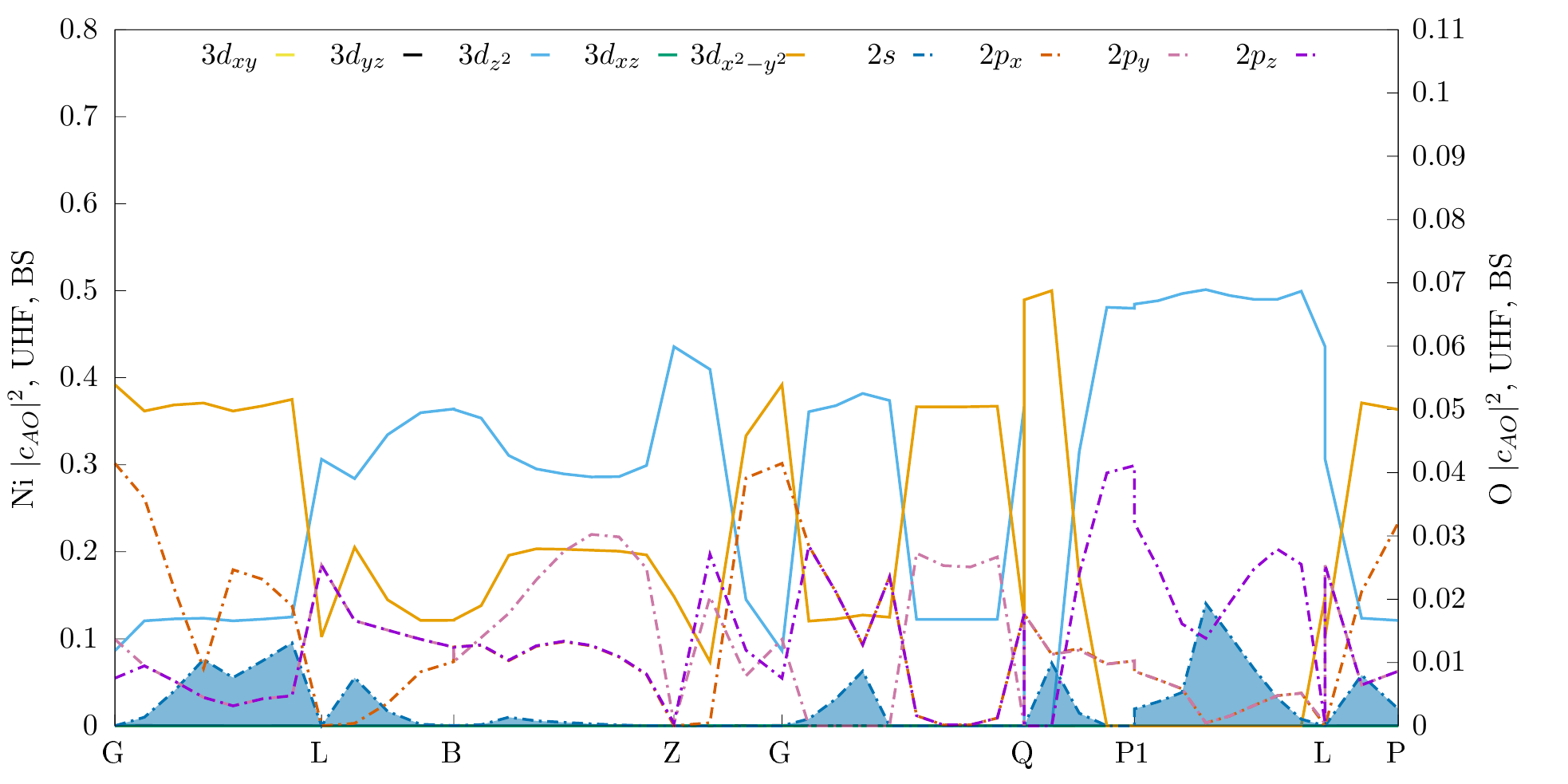} \\
  \includegraphics[width=8cm]{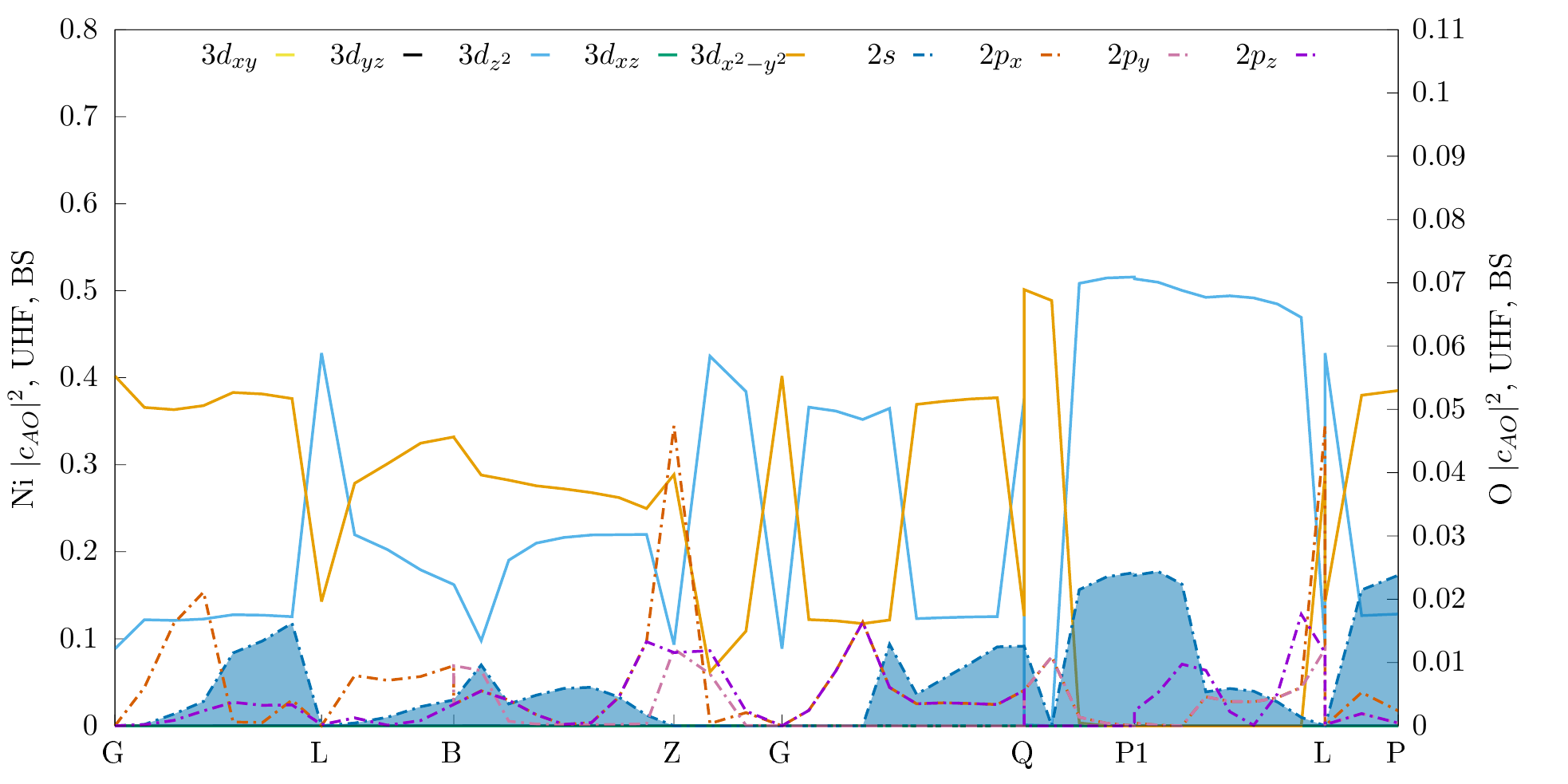}
\hfill
  \includegraphics[width=8cm]{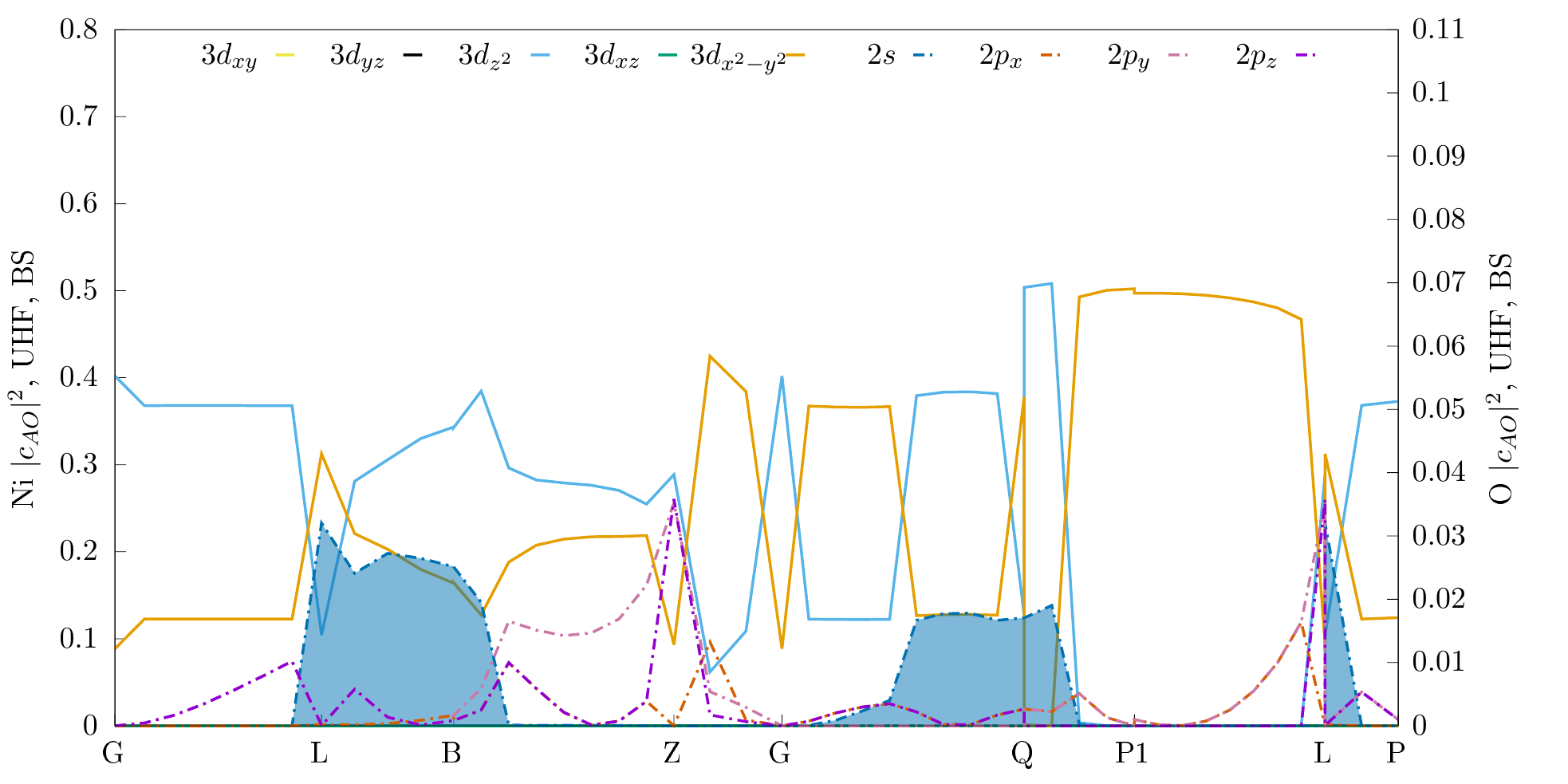} 
\centering
\caption{Squared absolute values of atomic contributions to the UHF frontier SA-NOs along Wannier interpolation path 
         in \textbf{cell 2}. 
         The contributions of $3d$ AOs from the first nickel atom in the cell are shown by solid curves; 
         the contributions of $2s$ and $2p$ AOs from the first oxygen atom are shown in dashed-dotted lines. 
         The areas below $2s$ curves are filled to highlight contributions from the $s$-orbitals. 
         \protect\label{fig:NiO_60_UHF_NO_AO}
}
\end{figure}
\begin{figure}[!h]
  \includegraphics[width=8cm]{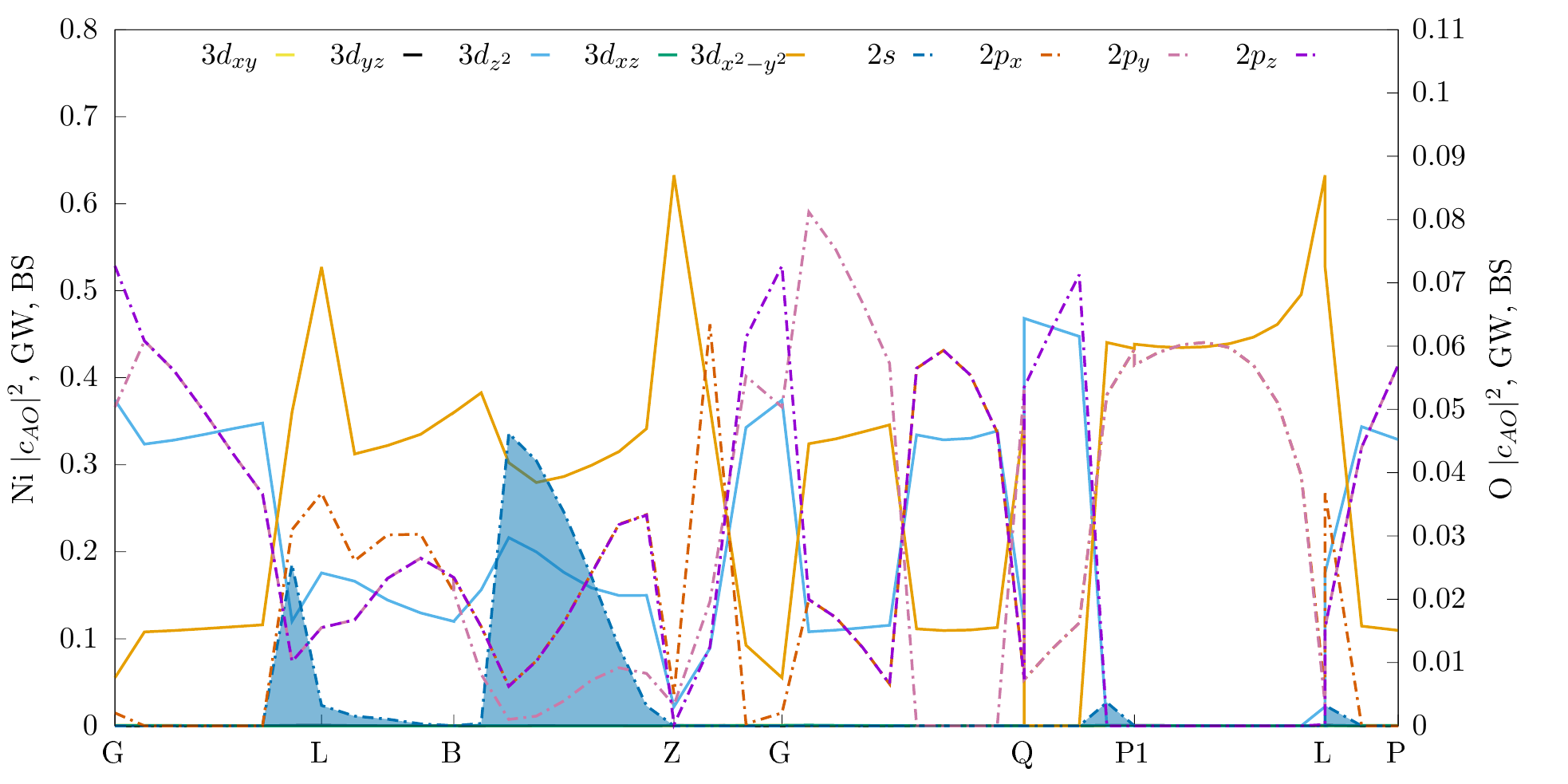}
\hfill
  \includegraphics[width=8cm]{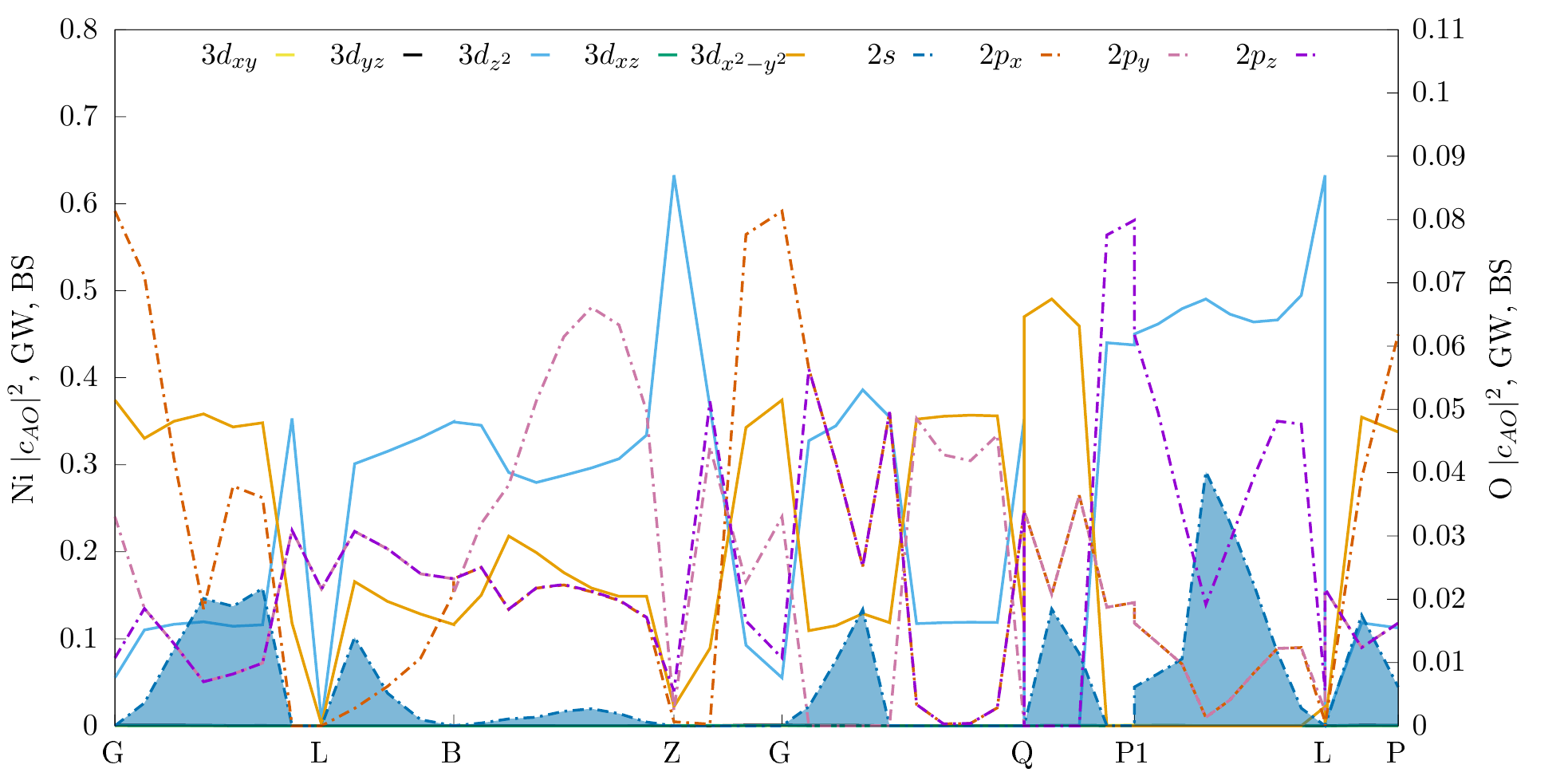} \\
  \includegraphics[width=8cm]{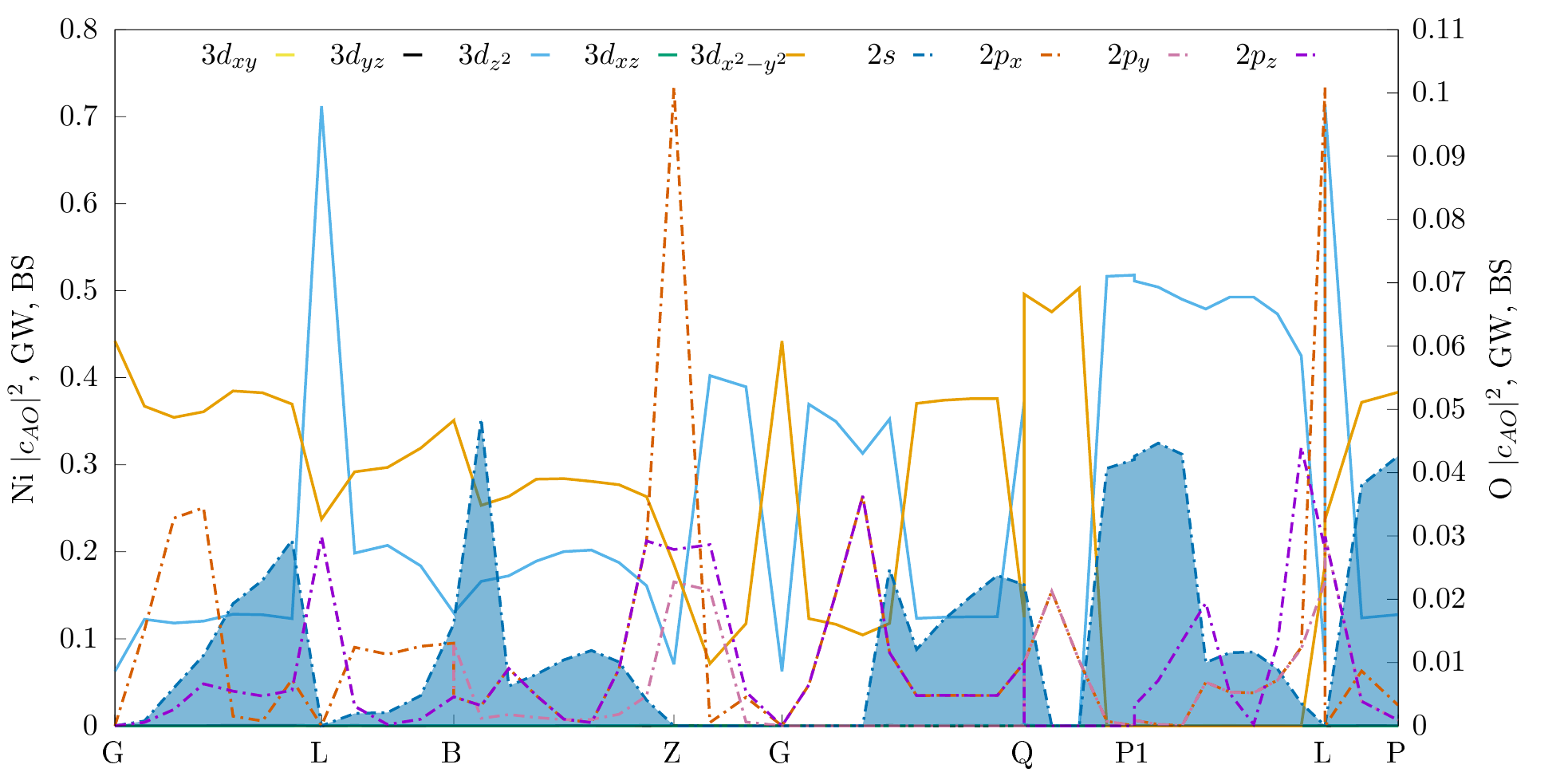}
\hfill
  \includegraphics[width=8cm]{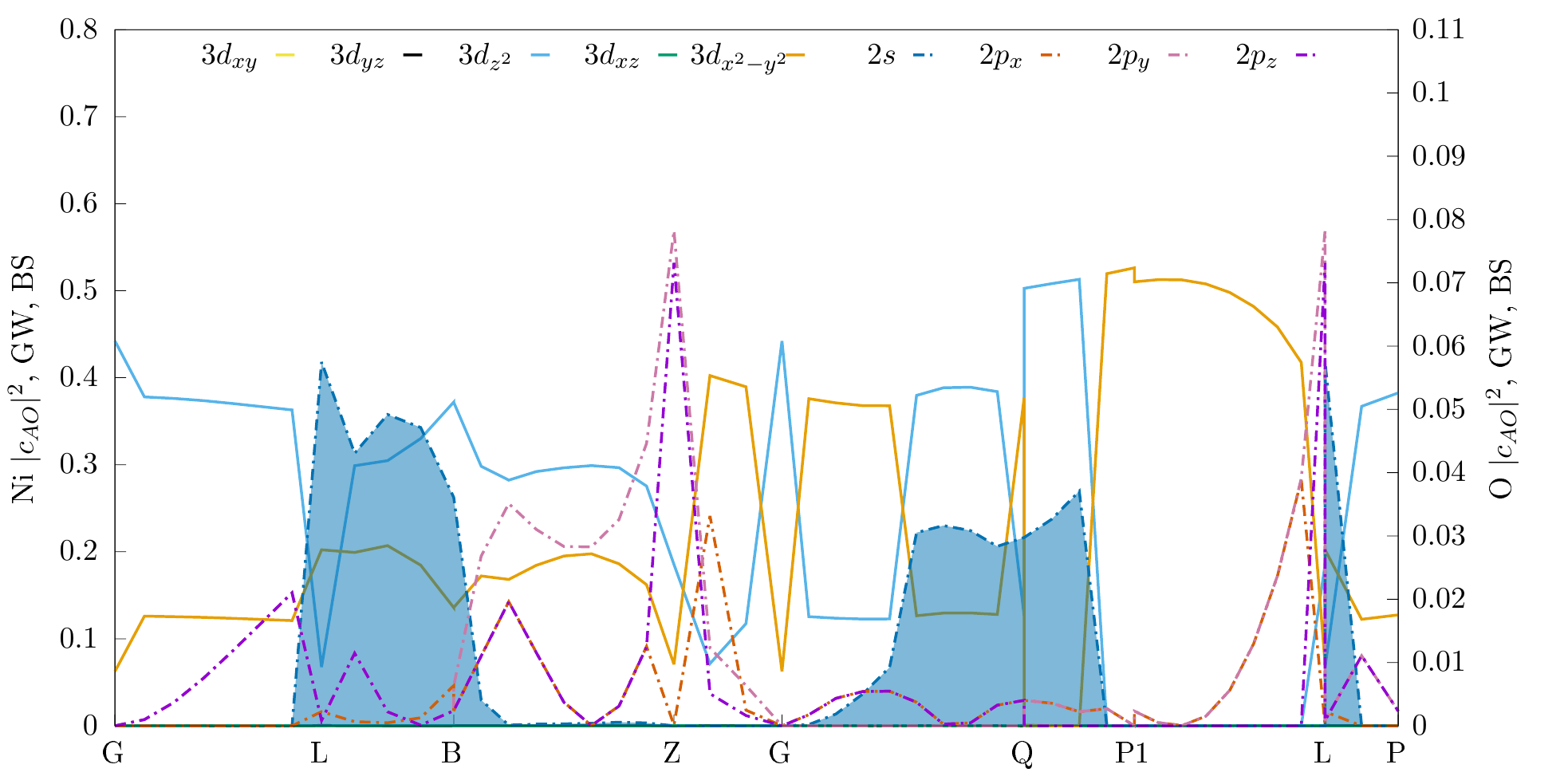} 
\centering
\caption{Squared absolute values of atomic contributions to the GW frontier SA-NOs along Wannier interpolation path 
         in \textbf{cell 2}. 
         The contributions of $3d$ AOs from the first nickel atom in the cell are shown by solid curves; 
         the contributions of $2s$ and $2p$ AOs from the first oxygen atom are shown in dashed-dotted lines. 
         The areas below $2s$ curves are filled to highlight contributions from the $s$-orbitals. 
         \protect\label{fig:NiO_60_GW_NO_AO}
}
\end{figure}

Figs.~\ref{fig:NiO_60_UHF_NO_AO} and \ref{fig:NiO_60_GW_NO_AO} show squares of AO coefficients in BS UHF and GW SA-NOs in \textbf{cell 2} for NiO. 
While the character of the frontier SA-NOs change significantly along the k-path, 
only $3d_{z^2}$ and $3d_{x^2-y^2}$ AOs on nickel contribute in a chosen coordinate system, 
while the weights of other d-orbitals are negligible. 
Contributions from the oxygen orbitals also vary along the k-path, violating GK rules. 
All the $2p$ orbitals contribute at different k-points. 
The contribution of the oxygen  $2s$ orbitals is very significant (highlighted by the blue fill) and is comparable with the $2p$ orbitals, which is also not expected from the GK rules. 
The inclusion of electron correlation with GW does not change the picture qualitatively, 
but the weights of all contributions from oxygen increase, while the weights from nickel $d$-orbitals decrease. 
All of these observations are also true for the BS solution in \textbf{cell 1} (Figs.~S2 and S3 in SI) along the corresponding k-path, 
where the weights of the $2s$ orbital on oxygen in SA-NOs are even bigger than in \textbf{cell 2}.  
Figs.~S6, S7, S8, and S9 in SI show the AO composition of the frontier SA-NOs for CoO in  \textbf{cell 2} and \textbf{cell 1}. 
Generally, all five $d$-orbitals contribute to SA-NOs, but in \textbf{cell 2} the most depopulated and the most populated frontier SA-NO have characters, similar to the ones for NiO. 
This suggests that the mechanism of superexchange leading to $J_2$ is similar in both compounds. 
The character of SA-NOs in \textbf{cell 1}  does not show this similarity, 
and the character of the most depopulated and populated frontier SA-NOs is mainly determined by cobalt $d_{xy}$, $d_{yz}$, and $d_{x^2-y^2}$  AOs, indicating a different mechanism leading to $J_1$.  
Figs.~S12, S13, S14, and S15 in SI show the AO composition of the frontier SA-NOs for FeO in  \textbf{cell 2} and \textbf{cell 1}. 
Again, all five $d$-orbitals contribute to SA-NOs, but in \textbf{cell 2} the most depopulated and the most populated frontier SA-NO have a character, mostly similar to the ones for NiO and CoO (there is a local intrusion of $d_{yz}$ and $d_{xy}$ AOs near $B$ and $Q$ points).  
In \textbf{cell 1}, the most depopulated and the most populated frontier SA-NOs are strongly dominated by $d_{yz}$ and are different from both NiO and CoO. 
In all the cells for all the considered compounds, GW preserves the overall character of the frontier SA-NOs, 
but increases the weights of the $s$- and $p$-orbitals on oxygen and decreases the weights of the $d$-orbitals on metal atoms.

\subsection{Two-particle correlators}
\begin{figure}[!h]
  \includegraphics[width=7cm]{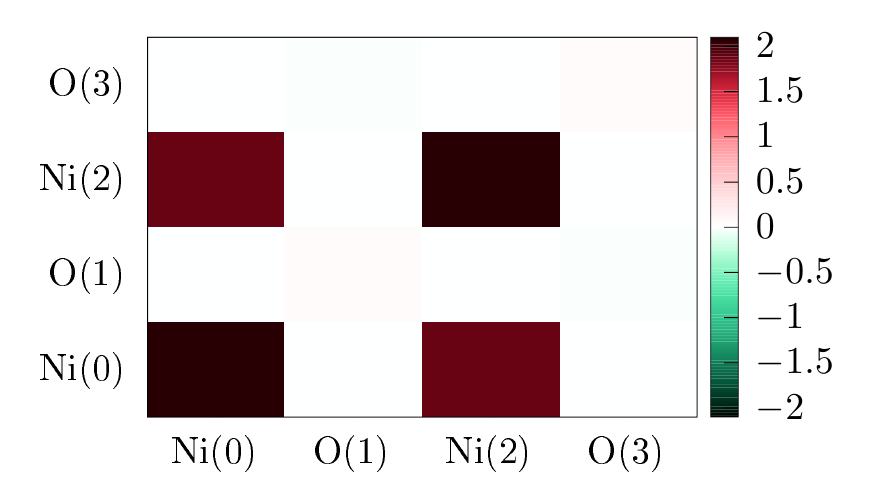}
\hfill
  \includegraphics[width=7cm]{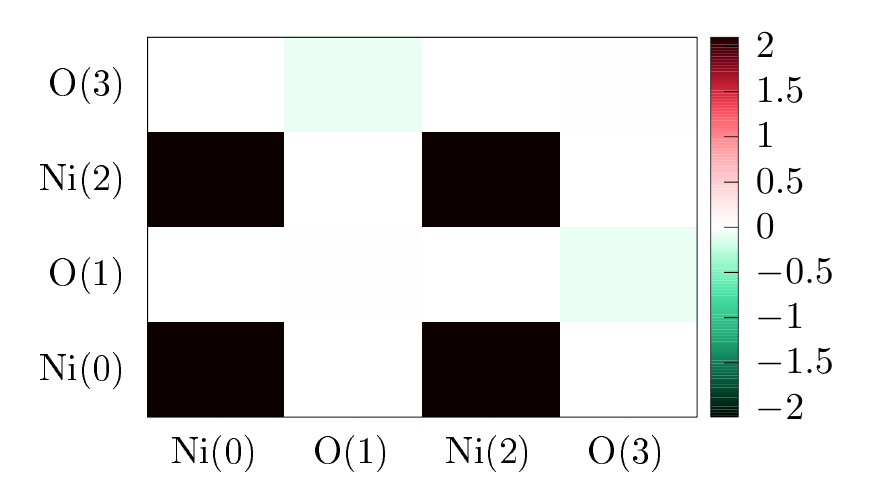}
\centering
\caption{Difference of $NN_{AB}^{\mathbf{k}=0}$ 
         between the HS and BS solutions, computed for the 
         NiO \textbf{cell 2} with UHF (left) and GW (right) with $5\times 5\times 5$ grid. 
         \protect\label{fig:NiO_60_NNloc_gamma}
}
\end{figure}
\begin{figure}[!h]
  \includegraphics[width=7cm]{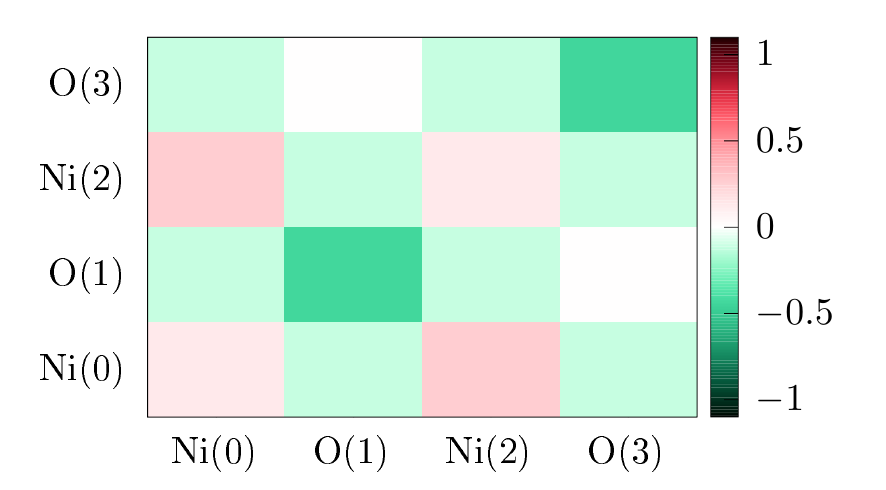}
\hfill
  \includegraphics[width=7cm]{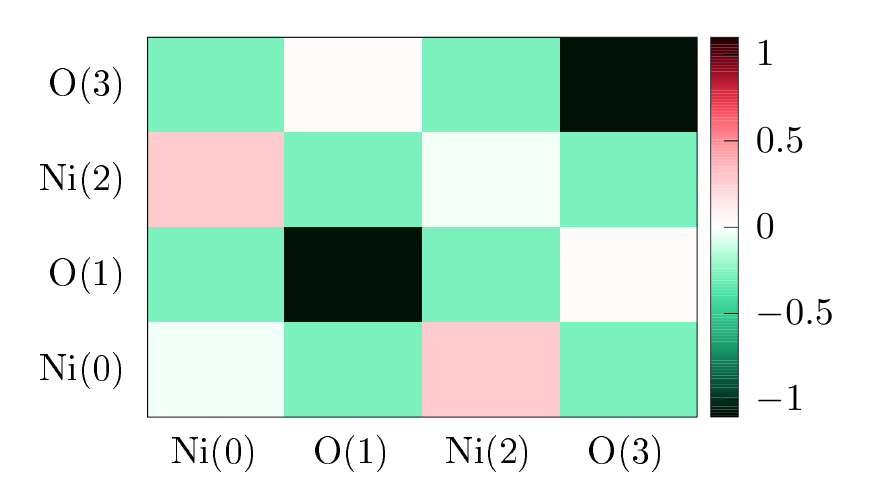}
\centering
\caption{Difference of $NN_{AB}$ 
         between the HS and BS solutions, computed for the 
         NiO \textbf{cell 2} with UHF (left) and GW (right) with $5\times 5\times 5$ grid. 
         \protect\label{fig:NiO_60_NNloc}
}
\end{figure}
The local two-particle correlators are shown in Figs. \ref{fig:NiO_60_NNloc_gamma} and \ref{fig:NiO_60_NNloc}. 
The $NN_{AB}^{\mathbf{k}=0}$ correlators show a significant difference in charge correlators on nickel centers, 
while the correlator difference between nickel and oxygen is surprisingly small. 
A possible cause behind this behaviour is due to smaller contributions from a single oxygen AOs to SA-NO \#1 and \#2 
in Fig.~\ref{fig:NiO_60_orbs} 
$\sum_i c^2_{2p_i} = 0.062$ (UHF) and 
$\sum_i c^2_{2p_i} = 0.125$ (GW) than from a single nickel AOs $\sum_j c^2_{3d_j} = 0.479$ (UHF) and $\sum_j c^2_{3d_j} = 0.431$ (GW), where we do not include weights of polarization functions for simplicity.  
The $NN_{AB}$ averages across all $k$-points, reducing the difference for Ni--Ni correlators and increasing the difference for the Ni--O correlators. 
Since the $NN_{AB}$ corresponds to a real-space charge correlator in localized orbitals, 
it clearly indicates that locally charge transfer from oxygen to nickel is rather substantial. 

\subsection{Role of electron correlation}
A renormalized interaction $W$, captured by GW, satisfies a Dyson-like equation.  
Because of this renormalization, $W$ describes screening that reduces the strength of the Coulomb interaction. 
Therefore, one would expect that the weight of charge-transfer contributions would become more prominent. 
In particular, one can expect an enhancement of charge transfer between oxygen and metal centers. 
This leads to an increase of AO coefficients of $s$- and $p$-orbitals and decrease of the coefficients of $d$-orbitals, 
which we observe for all SA-NOs for all the compounds in all the cells. 
Screening also leads to an increase of occupancy deviation from 1 for the broken-symmetry solutions for 
all the compounds and for all the cells, while preserving the overall character of the SA-NOs. 
At the two-particle level, screening also enhances the differences in local charge correlators, 
which also confirms the increase of the charge-transfer character. 

\section{Conclusions and implications for the future}
We generalized the concepts of natural orbitals and spin-averaged natural orbitals. 
We also generalized the $n_{l}$ and $n_{nl}$ indices---the effective numbers of open-shell electrons---for solids. 
Using these concepts, we analyzed electronic structure of 
cubic rock-salt transition-metal oxides: NiO, CoO, FeO. 
Our findings disagree with the qualitative Goodenough--Kanamori rules, formulated on semiempirical grounds almost half a century ago.  
Only the frontier SA-NOs are responsible for magnetic physics. 
Specifically, a complicated momentum behavior of the SA-NO occupation numbers for the broken-symmetry solutions is in a disagreement with the predictions deduced from Goodenough--Kanamori rules. 
All the frontier SA-NOs of the broken-symmetry solutions form pairs, net occupancy of which is close to 2; 
occupancies of all the frontier SA-NOs of the high-spin solutions is close to 1. 
This gives a clear picture of the effective exchange coupling, 
related to a chemical concept of multielectron multicenter bonding.   
The character of the frontier SA-NOs of the broken-symmetry solutions change significantly along the interpolated k-paths in the corresponding unit cells. 
This is also in a disagreement with Goodenough--Kanamori rules. 
In particular, we found that $p$-orbitals on oxygen are as important as $s$-orbitals, 
which have not been considered before in simplistic models.  
This finding not only affects the qualitative picture of magnetic interactions, 
but also indicates that $s$-orbitals must be included in the active spaces used in embedded methods, 
such as a self-energy embedding theory\cite{Kananenka15,Tran_GW_SEET,Tran_jcp_2015,Tran16,Tran_useet,Zgid17,Rusakov:SEET:2019,Iskakov20,Yeh:perovskite:2021}. 
The character and physics of the SA-NOs in our calculations are qualitatively different from the NOs 
of the localized clusters, considered in Ref.\citenum{deGraaf:CaMnO3:charge_transfer:2007}.  
The effective number of open-shell electrons, captured by $n_{l,\mathbf{k}}$ and $n_{nl,\mathbf{k}}$ indices, 
show how dynamic correlation changes the occupation numbers. 

We found that in the considered insulators the analysis of the frontier SA-NOs provides more insights into 
the mechanism of effective exchange couplings.  
We also generalized local two-particle charge correlators and used them to confirm the changes in charge-transfer character of the solutions. 
Inclusion of correlation through screened interaction $W$ enhanced the charge-transfer character of the solutions, 
which is seen in SA-NO occupancies, SA-NO compositions, and two-particle correlators. 
We and others previously observed the enhancement of charge-transfer character by dynamic correlation in multiple bridged molecular transition-metal complexes\cite{Pokhilko:local_correlators:2021,Morokuma:DMRG:biquad_exc:2014,Gagliardi:Cr2muOH:superexchange:2020}; 
however, explanations of this phenomenon in the multireference wave-function methods was lacking, 
which we now can attribute to the presence of screening.  
Broken-symmetry fully self-consistent GW gives much more accurate values of effective exchange couplings. 
This analysis implies, for example, that a complete-active-space self-consistent field method should give poor estimates of the effective exchange couplings due to insufficient screening, explaining why only very large active spaces are sufficient for reasonable estimates, which is often observed in practice.  
The analysis above also has implications for DFT. 
While the majority of DFT functionals give unsatisfactory effective exchange constants for transition-metal oxides, 
the best-performing functional from Ref\cite{Majumdar:NiO:MnO:DFT:J:2011} is HSE. 
This functional is based on Coulomb attenuation of PBE in the long-range part with some empirical mixture of HF exchange and PBE in the short-range part\cite{HSE:2003}. 
The Coulomb attenuation mimics the effects of screening, which is the most important part of the magnetic physics in these systems, explaining a good performance of this functional in comparison to other functionals without screening.  
Due to this theoretical justification, we recommend to start search of DFT functional for evaluation of effective exchange couplings from the Coulomb-attenuated family of methods.  
Finally, since broken-symmetry fully self-consistent GW gives  results comparable with multireference CI\cite{Pokhilko:local_correlators:2021} and difference-dedicated CI\cite{Pokhilko:BS-GW:solids:2022}, 
it is a good starting point for a perturbation theory based on renormalized interaction $W$ 
and renormalized propagator $G$ and included through a 3-point vertex function. 
We predict that such a screened perturbation theory converges rapidly and 
gives highly accurate effective exchange couplings, which we will pursue in the near future.

\section*{Acknowledgments}
P.P. and D.Z. acknowledge support from 
were supported by the Simons Foundation via the Simons Collaboration
on the Many-Electron problem.
We thank Dr. Gaurav Harsha for providing a band plotting script, 
which we modified to produce the plots along the interpolated $k$-paths.

\section*{Supplementary Material}
NiO natural orbitals; atomic contributions to the frontier SA-NOs for NiO, CoO, and FeO; 
occupancies of the frontier SA-NOs for CoO and FeO. 


\renewcommand{\baselinestretch}{1.5}


\begin{thebibliography}{100}

\bibitem{Malrieu:DDCI:1993}
J.~Miralles, O.~Castell, R.~Caballol, and J.-P. Malrieu,
\newblock Specific {CI} calculation of energy differences: Transition energies
  and bond energies,
\newblock Chem. Phys. {\bf 172}, 33 (1993).

\bibitem{Zimmerman:iFCI:exchange:2021}
A.~E. Rask and P.~M. Zimmerman,
\newblock Toward full configuration interaction for transition-metal complexes,
\newblock J. Phys. Chem. A {\bf 125}, 1598 (2021).

\bibitem{Roos:90:CASPT2}
K.~Andersson, P.-{\AA}. Malmqvist, B.~O. Roos, A.J. Sadlej, and K.~Wolinski,
\newblock 2nd order perturbation theory with a {C}{A}{S}{S}{C}{F} reference
  function,
\newblock J. Phys. Chem. {\bf 94}, 5483 (1990).

\bibitem{Angeli1}
Celestino Angeli, Renzo Cimiraglia, and Jean-Paul Malrieu,
\newblock n-electron valence state perturbation theory: A spinless formulation
  and an efficient implementation of the strongly contracted and of the
  partially contracted variants,
\newblock J. Chem. Phys. {\bf 117}, 9138 (2002).

\bibitem{Angeli2}
C.~Angeli, R.~Cimiraglia, S.~Evangelisti, T.~Leininger, and J.-P. Malrieu,
\newblock Introduction of n-electron valence states for multireference
  perturbation theory,
\newblock J. Chem. Phys. {\bf 114}, 10252 (2001).

\bibitem{Casanova:SFReview}
D.~Casanova and A.~I. Krylov,
\newblock Spin-flip methods in quantum chemistry,
\newblock Phys. Chem. Chem. Phys. {\bf 22}, 4326 (2020).

\bibitem{Krylov:SFTDDFT:2003}
Y.~Shao, M.~Head-Gordon, and A.~I. Krylov,
\newblock The spin-flip approach within time-dependent density functional
  theory: Theory and applications to diradicals,
\newblock J. Chem. Phys. {\bf 118}, 4807 (2003).

\bibitem{Bernard:SF:12}
Y.~A. Bernard, Y.~Shao, and A.~I. Krylov,
\newblock General formulation of spin-flip time-dependent density functional
  theory using non-collinear kernels: Theory, implementation, and benchmarks,
\newblock J. Chem. Phys. {\bf 136}, 204103 (2012).

\bibitem{Ziegler:sfdft:04}
F.~Wang and T.~Ziegler,
\newblock Time-dependent density functional theory based on a noncollinear
  formulation of the exchange-correlation potential,
\newblock J. Chem. Phys. {\bf 121}, 12191 (2004).

\bibitem{Ziegler:sa:cvdft:11}
H.~R. Zhekova, M.~Seth, and T.~Ziegler,
\newblock Calculation of the exchange coupling constants of copper binuclear
  systems based on spin-flip constricted variational density functional theory,
\newblock J. Chem. Phys. {\bf 135}, 184105 (2011).

\bibitem{Ziegler:Cu2:2011}
H.~R. Zhekova, M.~Seth, and T.~Ziegler,
\newblock Calculation of the exchange coupling constants of copper binuclear
  systems based on spin-flip constricted variational density functional theory,
\newblock J. Chem. Phys. {\bf 135}, 184105 (2012).

\bibitem{Ziegler:Cu2tb:2012}
I.~Seidu, H.~R. Zhekova, M.~Seth, and T.~Ziegler,
\newblock Calculation of exchange coupling constants in triply-bridged
  dinuclear {Cu(II)} compounds based on spin-flip constricted variational
  density functional theory,
\newblock J. Phys. Chem. A {\bf 116}, 2268 (2012).

\bibitem{Ziegler:Cu3:11}
H.~R. Zhekova, M.~Seth, and T.~Ziegler,
\newblock Introduction of a new theory for the calculation of magnetic coupling
  based on spin flip constricted variational density functional theory.
  {Application} to trinuclear copper complexes which model the native
  intermediate in multicopper oxidases,
\newblock J. Chem. Theory Comput. {\bf 7}, 1858–1866 (2011).

\bibitem{Liu:SFtensor:10}
Z.~D. Li and W.~J. Liu,
\newblock Spin-adapted open-shell random phase approximation and time-dependent
  density functional theory. {I. Theory},
\newblock J. Chem. Phys. {\bf 133}, 064106 (2010).

\bibitem{Liu:SFtensor:11}
Z.~D. Li, W.~J. Liu, Y.~Zhang, and B.~B. Suo,
\newblock Spin-adapted open-shell time-dependent density functional theory.
  {II. Theory} and pilot application,
\newblock J. Chem. Phys. {\bf 134}, 134101 (2011).

\bibitem{Valero:SFDFT:11}
R.~Valero, F.~Illas, and D.~G. Truhlar,
\newblock Magnetic coupling in transition-metal binuclear complexes by
  spin-flip time-dependent density functional theory,
\newblock J. Chem. Theory Comput. {\bf 7}, 3523 (2011).

\bibitem{Orms:magnets:17}
N.~Orms and A.~I. Krylov,
\newblock Singlet-triplet energy gaps and the degree of diradical character in
  binuclear copper molecular magnets characterized by spin-flip density
  functional theory,
\newblock Phys. Chem. Chem. Phys. {\bf 20}, 13127 (2018).

\bibitem{Kotaru:Fe:SMM:2022}
S.~Kotaru, S.~Kähler, M.~Alessio, and A.~I. Krylov,
\newblock Magnetic exchange interactions in binuclear and tetranuclear
  iron({III}) complexes described by spin-flip {DFT} and {H}eisenberg effective
  {H}amiltonians,
\newblock J. Comput. Chem. {\bf n/a}.

\bibitem{sfpaper}
A.~I. Krylov,
\newblock Size-consistent wave functions for bond-breaking: The
  equation-of-motion spin-flip model,
\newblock Chem. Phys. Lett. {\bf 338}, 375 (2001).

\bibitem{Casanova:2SF:08}
D.~Casanova, L.~V. Slipchenko, A.~I. Krylov, and M.~Head-Gordon,
\newblock Double spin-flip approach within equation-of-motion coupled cluster
  and configuration interaction formalisms: Theory, implementation and
  examples,
\newblock J. Chem. Phys. {\bf 130}, 044103 (2009).

\bibitem{Jagau:SF-CC2:2022}
G.~P. Paran, C.~Utku, and T.-C. Jagau,
\newblock A spin-flip variant of the second-order approximate coupled-cluster
  singles and doubles method,
\newblock Phys. Chem. Chem. Phys. {\bf 24}, 27146 (2022).

\bibitem{Yamaguchi:APUMP:1989}
Y.~Takahara, K.~Yamaguchi, and T.~Fueno,
\newblock Potential energy curves for transition metal dimers and complexes
  calculated by the approximately projected unrestricted {H}artree--{F}ock and
  {M}{\o}ller--{P}lesset perturbation ({APUMP}) methods,
\newblock Chem. Phys. Lett. {\bf 158}, 95 (1989).

\bibitem{Yamaguchi:APCCSD:2012}
T.~Saito, A.~Ito, T.~Watanabe, T.~Kawakami, M.~Okumura, and K.~Yamaguchi,
\newblock Performance of the coupled cluster and {DFT} methods for
  through-space magnetic interactions of nitroxide dimer,
\newblock Chem. Phys. Lett. {\bf 542}, 19 (2012).

\bibitem{Stanton:BS-CC:2020}
H.~Schurkus, D.-T. Chen, H.-P. Cheng, G.~Chan, and J.~Stanton,
\newblock Theoretical prediction of magnetic exchange coupling constants from
  broken-symmetry coupled cluster calculations,
\newblock J. Chem. Phys. {\bf 152}, 234115 (2020).

\bibitem{Chibotaru:BS-G0W0:2020}
A.~Mansikkam{\"a}ki, Z.~Huang, N.~Iwahara, and L.~F. Chibotaru,
\newblock Broken symmetry {$G_0 W_0$} approach for the evaluation of exchange
  coupling constants,
\newblock https://arxiv.org/abs/2003.06334  (2020).

\bibitem{Pokhilko:local_correlators:2021}
P.~Pokhilko and D.~Zgid,
\newblock Interpretation of multiple solutions in fully iterative {GF2} and
  {GW} schemes using local analysis of two-particle density matrices,
\newblock J. Chem. Phys. {\bf 155}, 024101 (2021).

\bibitem{Pokhilko:BS-GW:solids:2022}
P.~Pokhilko and D.~Zgid,
\newblock Broken-symmetry self-consistent {GW} approach: Degree of spin
  contamination and evaluation of effective exchange couplings in solid
  antiferromagnets,
\newblock J. Chem. Phys. {\bf 157}, 144101 (2022).

\bibitem{Liechtenstein:mag_force:1987}
A.I. Liechtenstein, M.I. Katsnelson, V.P. Antropov, and V.A. Gubanov,
\newblock Local spin density functional approach to the theory of exchange
  interactions in ferromagnetic metals and alloys,
\newblock J. Magn. Magn. Mater. {\bf 67}, 65 (1987).

\bibitem{Peralta:mag_force:2022}
L.~E. Aebersold, A.~R. Hale, G.~Christou, and J.~E. Peralta,
\newblock Validation of the {G}reen’s function approximation for the
  calculation of magnetic exchange couplings,
\newblock J. Phys. Chem. A {\bf 126}, 6790 (2022).

\bibitem{Martin:NiO:exchange:2002}
I.~de~P.~R.~Moreira, F.~Illas, and R.~L. Martin,
\newblock Effect of {F}ock exchange on the electronic structure and magnetic
  coupling in {NiO},
\newblock Phys. Rev. B {\bf 65}, 155102 (2002).

\bibitem{deGraaf:MRPT:exchange:solids:SMM:2001}
C.~de~Graaf, C.~Sousa, I.~de~P.~R.~Moreira, and F.~Illas,
\newblock Multiconfigurational perturbation theory: An efficient tool to
  predict magnetic coupling parameters in biradicals, molecular complexes, and
  ionic insulators,
\newblock J. Phys. Chem. A {\bf 105}, 11371 (2001).

\bibitem{Pokhilko:spinchain}
P.~Pokhilko, D.~S. Bezrukov, and A.~I. Krylov,
\newblock Is solid copper oxalate a spin chain or a mixture of entangled spin
  pairs?,
\newblock J. Phys. Chem. C {\bf 125}, 7502 (2021).

\bibitem{noodleman:BS:81}
L.~Noodleman,
\newblock Valence bond description of antiferromagnetic coupling in transition
  metal dimers,
\newblock J. Chem. Phys. {\bf 74}, 5737 (1981).

\bibitem{Yamaguchi:BS:formulation:1986}
K.~Yamaguchi, Y.~Takahara, and T.~Fueno,
\newblock Ab-initio molecular orbital studies of structure and reactivity of
  transition metal-oxo compounds,
\newblock in {\em Applied quantum chemistry}, pages 155--184. Springer, 1986.

\bibitem{Malrieu:spin_pol:BS-DFT:2020}
G.~David, N.~Ferr{\'e}, G.~Trinquier, and J.-P. Malrieu,
\newblock Improved evaluation of spin-polarization energy contributions using
  broken-symmetry calculations,
\newblock J. Chem. Phys. {\bf 153}, 054120 (2020).

\bibitem{Malrieu:decont:BS-DFT:2020}
G.~David, G.~Trinquier, and J.-P. Malrieu,
\newblock Consistent spin decontamination of broken-symmetry calculations of
  diradicals,
\newblock J. Chem. Phys. {\bf 153}, 194107 (2020).

\bibitem{Cremer:DFT:S2:2001}
J.~Gr{\"a}fenstein and D.~Cremer,
\newblock On the diagnostic value of ($\hat{S^2}$) in {K}ohn--{S}ham density
  functional theory,
\newblock Mol. Phys. {\bf 99}, 981 (2001).

\bibitem{Handy:DFT:S2:2007}
A.~J. Cohen, D.~J. Tozer, and N.~C. Handy,
\newblock Evaluation of $\braket{S^2}$ in density functional theory,
\newblock J. Chem. Phys. {\bf 126}, 214104 (2007).

\bibitem{Vedene:DFT:S2:1995}
J.~Wang, A.~D. Becke, and V.~H. Smith,
\newblock Evaluation of $\braket{S^2}$ in restricted, unrestricted
  {H}artree--{F}ock, and density functional based theories,
\newblock J. Chem. Phys. {\bf 102}, 3477 (1995).

\bibitem{Yamaguchi:APDFT:solid:2021}
K.~Tada, S.~Yamanaka, T.~Kawakami, Y.~Kitagawa, M.~Okumura, K.~Yamaguchi, and
  S.~Tanaka,
\newblock Estimation of spin contamination errors in {DFT}/plane-wave
  calculations of solid materials using approximate spin projection scheme,
\newblock Chem. Phys. Lett. {\bf 765}, 138291 (2021).

\bibitem{Mahan00}
G.~D. Mahan,
\newblock {\em Many-Particle Physics},
\newblock Physics of Solids and Liquids. Springer, 2000.

\bibitem{Negele:Orland:book:2018}
J.~W. Negele and H.~Orland,
\newblock {\em Quantum many-particle systems}. CRC Press, 2018.

\bibitem{Martin:Interacting_electrons:2016}
R.~M. Martin, L.~Reining, and D.~M. Ceperley,
\newblock {\em Interacting electrons}. Cambridge University Press, 2016.

\bibitem{Almbladh:photoemission:1985}
C.-O. Almbladh,
\newblock On the theory of photoemission,
\newblock Phys. Scr. {\bf 32}, 341 (1985).

\bibitem{Hedin:photoemission:1985}
W.~Bardyszewski and L.~Hedin,
\newblock A new approach to the theory of photoemission from solids,
\newblock Phys. Scr. {\bf 32}, 439 (1985).

\bibitem{Fujikawa:photoelectron:chapter:2015}
T.~Fujikawa and K.~Niki,
\newblock {\em Theory of Photoelectron Spectroscopy}, volume 209, pages
  285--301. Springer Japan, Tokyo, 2015.

\bibitem{Kadanoff:superconductivity:1961}
L.~P. Kadanoff and P.~C. Martin,
\newblock Theory of many-particle systems. {II}. {S}uperconductivity,
\newblock Phys. Rev. {\bf 124}, 670 (1961).

\bibitem{Luttinger60}
J.~M. Luttinger and J.~C. Ward,
\newblock Ground-state energy of a many-fermion system. {II},
\newblock Phys. Rev. {\bf 118}, 1417 (1960).

\bibitem{Baym61}
G.~Baym and L.~P. Kadanoff,
\newblock Conservation laws and correlation functions,
\newblock Phys. Rev. {\bf 124}, 287 (1961).

\bibitem{Baym62}
G.~Baym,
\newblock Self-consistent approximations in many-body systems,
\newblock Phys. Rev. {\bf 127}, 1391 (1962).

\bibitem{Pokhilko:tpdm:2021}
P.~Pokhilko, S.~Iskakov, C.-N. Yeh, and D.~Zgid,
\newblock Evaluation of two-particle properties within finite-temperature
  self-consistent one-particle {G}reen’s function methods: Theory and
  application to {GW} and {GF2},
\newblock J. Chem. Phys. {\bf 155}, 024119 (2021).

\bibitem{Cloizeax:1960}
J.~des Cloizeaux,
\newblock Extension d'une formule de lagrange {\`a} des probl{\`e}mes de
  valeurs propres,
\newblock Nucl. Phys. {\bf 20}, 321 (1960).

\bibitem{Bloch:1958}
C.~Bloch,
\newblock Sur la th{\`e}orie des perturbations des {\'e}tats li{\'e}s,
\newblock Nucl. Phys. {\bf 6}, 329 (1958).

\bibitem{Okubo:1954}
S.~{\^O}kubo,
\newblock Diagonalization of {H}amiltonian and {T}amm-{D}ancoff equation,
\newblock Prog. Theor. Phys. {\bf 12}, 603 (1954).

\bibitem{Durand:EffHam:1983}
P.~Durand,
\newblock Direct determination of effective hamiltonians by wave-operator
  methods. {I. General} formalism,
\newblock Phys. Rev. A {\bf 28}, 3184 (1983).

\bibitem{Soliverez:1969}
C.~E. Soliverez,
\newblock An effective hamiltonian and time-independent perturbation theory,
\newblock J. Phys. C: Solid State Phys. {\bf 2}, 2161 (1969).

\bibitem{Calzado:02}
C.~J. Calzado, J.~Cabrero, J.~P. Malrieu, and R.~Caballol,
\newblock Analysis of the magnetic coupling in binuclear complexes. {II}.
  {D}erivation of valence effective hamiltonians from ab initio {CI} and {DFT}
  calculations,
\newblock J. Chem. Phys. {\bf 116}, 3985 (2002).

\bibitem{Marlieu:MagnetRev:2014}
J.~P. Malrieu, R.~Caballol, C.~J. Calzado, C.~{de~Graaf}, and N.~Guih{\'e}ry,
\newblock Magnetic interactions in molecules and highly correlated materials:
  {Physical} content, analytical derivation, and rigorous extraction of
  magnetic {Hamiltonians},
\newblock Chem. Rev. {\bf 114}, 429 (2013).

\bibitem{Buchachenko:Mn2:2010}
A.~A. Buchachenko, G.~Cha{\l{}}asi{\'n}ski, and M.~M. Szcz{\k{e}}{\'s}niak,
\newblock Electronic structure and spin coupling of the manganese dimer: The
  state of the art of ab initio approach,
\newblock J. Chem. Phys. {\bf 132}, 024312 (2010).

\bibitem{Mayhall:2014:HDVV}
N.~J. Mayhall and M.~Head-Gordon,
\newblock Computational quantum chemistry for single {Heisenberg} spin
  couplings made simple: {Just} one spin flip required,
\newblock J. Chem. Phys. {\bf 141}, 134111 (2014).

\bibitem{Mayhall:1SF:2015}
N.~J. Mayhall and M.~Head-Gordon,
\newblock Computational quantum chemistry for multiple-site {Heisenberg} spin
  couplings made simple: {Still} only one spin-flip required,
\newblock J. Phys. Chem. Lett. {\bf 6}, 1982 (2015).

\bibitem{Pokhilko:EffH:2020}
P.~Pokhilko and A.~I. Krylov,
\newblock Effective {H}amiltonians derived from equation-of-motion
  coupled-cluster wave-functions: {T}heory and application to the {H}ubbard and
  {H}eisenberg {H}amiltonians,
\newblock J. Chem. Phys. {\bf 152}, 094108 (2020).

\bibitem{Pokhilko:Neel_T:2022}
P.~Pokhilko and D.~Zgid,
\newblock Evaluation of {N}eel temperatures from fully self-consistent
  broken-symmetry {GW} and high-temperature expansion: application to cubic
  transition-metal oxides,
\newblock page https://arxiv.org/abs/2209.14904 (2022).

\bibitem{Kramers:superexchange:1934}
H.~A. Kramers,
\newblock L'interaction entre les atomes magnétogènes dans un cristal
  paramagnétique,
\newblock Physica {\bf 1}, 182 (1934).

\bibitem{Anderson:superexchange:1950}
P.~W. Anderson,
\newblock Antiferromagnetism. theory of superexchange interaction,
\newblock Phys. Rev. {\bf 79}, 350 (1950).

\bibitem{Anderson:exchange:1963}
P.~W. Anderson,
\newblock Theory of magnetic exchange interactions:exchange in insulators and
  semiconductors,
\newblock volume~14 of {\em Solid State Phys.}, pages 99--214. Academic Press,
  1963.

\bibitem{Eremin:exchange:1980}
M.~V. Eremin and Yu.~V. Rakitin,
\newblock On kinetic exchange theory,
\newblock Phys. Stat. Sol. (b) {\bf 97}, 51 (1980).

\bibitem{Coulaud:Jdecomp:2012}
E.~Coulaud, N.~Guihéry, J.-P. Malrieu, D.~Hagebaum-Reignier, D.~Siri, and
  N.~Ferré,
\newblock Analysis of the physical contributions to magnetic couplings in
  broken symmetry density functional theory approach,
\newblock J. Chem. Phys. {\bf 137}, 114106 (2012).

\bibitem{Coulaud:Jdecomp:2013}
E.~Coulaud, J.-P. Malrieu, N.~Guihéry, and N.~Ferré,
\newblock Additive decomposition of the physical components of the magnetic
  coupling from broken symmetry density functional theory calculations,
\newblock J. Chem. Theory Comput. {\bf 9}, 3429 (2013).

\bibitem{Ferre:Jdecomp:2018}
G.~David, F.~Wennmohs, F.~Neese, and N.~Ferré,
\newblock Chemical tuning of magnetic exchange couplings using broken-symmetry
  density functional theory,
\newblock Inorg. Chem. {\bf 57}, 12769 (2018).

\bibitem{LeGuennic:multicenter:Jdecomp:2023}
G.~David, N.~Ferré, and B.~Le~Guennic,
\newblock Consistent evaluation of magnetic exchange couplings in multicenter
  compounds in {KS-DFT}: The recomposition method,
\newblock J. Chem. Theory Comput. {\bf 19}, 157 (2023).

\bibitem{note:time}
In principle, one can merge momentum, spin, space coordinate, and time. For
  example, the corresponding multiindex is a convenient label for two-particle
  Green's functions. However, in this paper, we focus on analysis of one- and
  two-particle quantities at zero time slice, so for our purposes we keep the
  time label separately.

\bibitem{note:spin_choice}
Note that other choices are possible as well with Green's functions, such as
  restricted ($G_{\alpha\alpha} = G_{\beta\beta}$, $G_{\alpha\beta} =
  G_{\beta\alpha} = 0$) and generalized ($G_{\alpha\alpha}$, $G_{\beta\beta}$,
  $G_{\alpha\beta}$, $G_{\beta\alpha}$ are non-zero and possibly different).

\bibitem{Luzanov:TDM-1:76}
A.~V. Luzanov, A.~A. Sukhorukov, and V.~E. Umanskii,
\newblock Application of transition density matrix for analysis of excited
  states,
\newblock Theor. Exp. Chem. {\bf 10}, 354 (1976),
\newblock Russian original: Teor. Eksp. Khim., 10, 456 (1974).

\bibitem{Luzanov:TDM-2:79}
A.~V. Luzanov and V.~F. Pedash,
\newblock Interpretation of excited states using charge-transfer number,
\newblock Theor. Exp. Chem. {\bf 15}, 338 (1979).

\bibitem{HeadGordon:att_det:95}
M.~Head-Gordon, A.~M. Grana, D.~Maurice, and C.~A. White,
\newblock Analysis of electronic transitions as the difference of electron
  attachment and detachment densities,
\newblock J. Phys. Chem. {\bf 99}, 14261  (1995).

\bibitem{Martin:NTO:03}
R.~L. Martin,
\newblock Natural transition orbitals,
\newblock J. Phys. Chem. A {\bf 118}, 4775 (2003).

\bibitem{Luzanov:DMRev:12}
A.~V. Luzanov and O.~A. Zhikol,
\newblock Excited state structural analysis: {TDDFT} and related models,
\newblock in {\em Practical aspects of computational chemistry {I}: {An}
  overview of the last two decades and current trends}, edited by
  J.~Leszczynski and M.K. Shukla, pages 415--449. Springer, 2012.

\bibitem{Dreuw:ESSAImpl:14}
F.~Plasser, M.~Wormit, and A.~Dreuw,
\newblock New tools for the systematic analysis and visualization of electronic
  excitations. {I}. formalism,
\newblock J. Chem. Phys. {\bf 141}, 024106 (2014).

\bibitem{Dreuw:ESSAImpl-2:14}
F.~Plasser, S.~A. B{\"a}ppler, M.~Wormit, and A.~Dreuw,
\newblock New tools for the systematic analysis and visualization of electronic
  excitations. {II. Applications},
\newblock J. Chem. Phys. {\bf 141}, 024107 (2014).

\bibitem{Plasser:excitons:2016}
S.~A. Mewes, J.-M. Mewes, A.~Dreuw, and F.~Plasser,
\newblock Excitons in poly(para phenylene vinylene): a quantum-chemical
  perspective based on high-level ab initio calculations,
\newblock Phys. Chem. Chem. Phys. {\bf 18}, 2548 (2016).

\bibitem{Nanda:NTO:17}
K.~D. Nanda and A.~I. Krylov,
\newblock Visualizing the contributions of virtual states to two-photon
  absorption cross-sections by natural transition orbitals of response
  transition density matrices,
\newblock J. Phys. Chem. Lett. {\bf 8}, 3256 (2017).

\bibitem{Wojtek:ImagEx:18}
W.~Skomorowski and A.~I. Krylov,
\newblock Real and imaginary excitons: Making sense of resonance wavefunctions
  by using reduced state and transition density matrices,
\newblock J. Phys. Chem. Lett. {\bf 9}, 4101 (2018).

\bibitem{Krylov:Libwfa:18}
S.~Mewes, F.~Plasser, A.~I. Krylov, and A.~Dreuw,
\newblock Benchmarking excited-state calculations using exciton properties,
\newblock J. Chem. Theory Comput. {\bf 14}, 710 (2018).

\bibitem{Dreuw:NTOfeature:2019}
S.~A. Mewes and A.~Dreuw,
\newblock Density-based descriptors and exciton analyses for visualizing and
  understanding the electronic structure of excited states,
\newblock Phys. Chem. Chem. Phys. {\bf 21}, 2843 (2019).

\bibitem{Pavel:SOCNTOs:2019}
P.~Pokhilko and A.~I. Krylov,
\newblock Quantitative {El-Sayed} rules for many-body wavefunctions from
  spinless transition density matrices,
\newblock J. Phys. Chem. Lett. {\bf 10}, 4857 (2019).

\bibitem{Krylov:Orbitals}
A.~I. Krylov,
\newblock From orbitals to observables and back,
\newblock J. Chem. Phys. {\bf 153}, 080901 (2020).

\bibitem{Plasser:Visualisation:2019}
F.~Plasser,
\newblock Visualisation of electronic excited-state correlation in real space,
\newblock ChemPhotoChem {\bf 3}, 702 (2019).

\bibitem{Nanda:RIXSNTO:20}
K.~D. Nanda and A.~I. Krylov,
\newblock A simple molecular orbital picture of {RIXS} distilled from many-body
  damped response theory,
\newblock J. Chem. Phys. {\bf 152}, 244118 (2020).

\bibitem{Wergifosse:respNTO:2020}
M.~de~Wergifosse and S.~Grimme,
\newblock A unified strategy for the chemically intuitive interpretation of
  molecular optical response properties,
\newblock J. Chem. Theory Comput. {\bf 16}, 7709 (2020).

\bibitem{Nanda:NTO:hyperpolarizability:2021}
K.~D. Nanda and A.~I. Krylov,
\newblock The orbital picture of the first dipole hyperpolarizability from
  many-body response theory,
\newblock J. Chem. Phys. {\bf 154}, 184109 (2021).

\bibitem{Kahn:book:1993}
O.~Kahn,
\newblock {\em Molecular Magnetism}. VCH, 1993.

\bibitem{Yamaguchi:magnetic:NO:2000}
T.~Soda, Y.~Kitagawa, T.~Onishi, Y.~Takano, Y.~Shigeta, H.~Nagao, Y.~Yoshioka,
  and K.~Yamaguchi,
\newblock Ab initio computations of effective exchange integrals for {H}--{H},
  {H}--{H}e--{H} and {M}n2{O}2 complex: comparison of broken-symmetry
  approaches,
\newblock Chem. Phys. Lett. {\bf 319}, 223 (2000).

\bibitem{Malrieu:mag_orbitals:2002}
J.~Cabrero, C.~J. Calzado, D.~Maynau, R.~Caballol, and J.~P. Malrieu,
\newblock Metal--ligand delocalization in magnetic orbitals of binuclear
  complexes,
\newblock J. Phys. Chem. A {\bf 106}, 8146 (2002).

\bibitem{Morokuma:DMRG:biquad_exc:2014}
T.~V. Harris, Y.~Kurashige, T.~Yanai, and K.~Morokuma,
\newblock Ab initio density matrix renormalization group study of magnetic
  coupling in dinuclear iron and chromium complexes,
\newblock J. Chem. Phys. {\bf 140}, 054303 (2014).

\bibitem{Gagliardi:Cr2muOH:superexchange:2020}
P.~Sharma, D.~G. Truhlar, and L.~Gagliardi,
\newblock Magnetic coupling in a tris-hydroxo-bridged chromium dimer occurs
  through ligand mediated superexchange in conjunction with through-space
  coupling,
\newblock J. Am. Chem. Soc. {\bf 142}, 16644 (2020).

\bibitem{Kresse:PW_NO:2011}
A.~Grüneis, G.~H. Booth, M.~Marsman, J.~Spencer, A.~Alavi, and G.~Kresse,
\newblock Natural orbitals for wave function based correlated calculations
  using a plane wave basis set,
\newblock J. Chem. Theory Comput. {\bf 7}, 2780 (2011).

\bibitem{Pucci:NOs_solids:2013}
N.~H. March, G.~G.~N. Angilella, and R.~Pucci,
\newblock Natural orbitals in relation to quantum information theory: from
  model light atoms through to emergent metallic properties,
\newblock Int. J. Mod. Phys. B {\bf 27}, 1330021 (2013).

\bibitem{Yamaguchi:index:1978}
K.~Takatsuka, T.~Fueno, and K.~Yamaguchi,
\newblock Distribution of odd electrons in ground-state molecules,
\newblock Theor. Chim. Acta {\bf 48}, 175 (1978).

\bibitem{Head-Gordon:Yamaguchi:03}
M.~Head-Gordon,
\newblock Characterizing unpaired electrons from the one-particle density
  matrix,
\newblock Chem. Phys. Lett. {\bf 372}, 508 (2003).

\bibitem{Hedin65}
L.~Hedin,
\newblock New method for calculating the one-particle {G}reen's function with
  application to the electron-gas problem,
\newblock Phys. Rev. {\bf 139}, A796 (1965).

\bibitem{G0W0_Pickett84}
W.~E. Pickett and C.~S. Wang,
\newblock Local-density approximation for dynamical correlation corrections to
  single-particle excitations in insulators,
\newblock Phys. Rev. B {\bf 30}, 4719 (1984).

\bibitem{G0W0_Hybertsen86}
M.~S. Hybertsen and S.~G. Louie,
\newblock Electron correlation in semiconductors and insulators: Band gaps and
  quasiparticle energies,
\newblock Phys. Rev. B {\bf 34}, 5390 (1986).

\bibitem{GW_Aryasetiawan98}
F.~Aryasetiawan and O.~Gunnarsson,
\newblock The {GW} method,
\newblock Rep. Prog. Phys. {\bf 61}, 237 (1998).

\bibitem{Stan06}
A.~Stan, N.~E. Dahlen, and R.~van Leeuwen,
\newblock Fully self-consistent {GW} calculations for atoms and molecules,
\newblock EPL {\bf 76}, 298 (2006).

\bibitem{Koval14}
P.~Koval, D.~Foerster, and D.~S\'anchez-Portal,
\newblock Fully self-consistent {GW} and quasiparticle self-consistent {GW} for
  molecules,
\newblock Phys. Rev. B {\bf 89}, 155417 (2014).

\bibitem{scGW_Andrey09}
A.~Kutepov, Sergey~Y. Savrasov, and G.~Kotliar,
\newblock Ground-state properties of simple elements from {GW} calculations,
\newblock Phys. Rev. B {\bf 80}, 041103(R) (2009).

\bibitem{Kutepov17}
A.~L. Kutepov,
\newblock Self-consistent solution of {H}edin's equations: Semiconductors and
  insulators,
\newblock Phys. Rev. B {\bf 95}, 195120 (2017).

\bibitem{Iskakov20}
S.~Iskakov, C.-N. Yeh, E.~Gull, and D.~Zgid,
\newblock Ab initio self-energy embedding for the photoemission spectra of
  {NiO} and {MnO},
\newblock Phys. Rev. B {\bf 102}, 085105 (2020).

\bibitem{Yeh:GPU:GW:2022}
C.-N. Yeh, S.~Iskakov, D.~Zgid, and E.~Gull,
\newblock Fully self-consistent finite-temperature {GW} in {G}aussian {B}loch
  orbitals for solids,
\newblock Phys. Rev. B {\bf 106}, 235104 (2022).

\bibitem{Kubo:cumulant:1962}
R.~Kubo,
\newblock Generalized cumulant expansion method,
\newblock J. Phys. Soc. Jap. {\bf 17}, 1100 (1962).

\bibitem{Ruedenberg:chem_bond:1962}
K.~Ruedenberg,
\newblock The physical nature of the chemical bond,
\newblock Rev. Mod. Phys. {\bf 34}, 326 (1962).

\bibitem{Jorge:bond_index:1985}
M.~S. de~Giambiagi, M.~Giambiagi, and F.~E. Jorge,
\newblock Bond index: relation to second-order density matrix and charge
  fluctuations,
\newblock Theor. Chim. Acta {\bf 68}, 337 (1985).

\bibitem{Torre:popul:cumulants:2002}
A.~Torre, L.~Lain, R.~Bochicchio, and R.~Ponec,
\newblock Topological population analysis from higher order densities {II}.
  {T}he correlated case,
\newblock J. Math. Chem. {\bf 32}, 241 (2002).

\bibitem{Bochicchio:bond_order:2003}
A.~Torre, L.~Lain, and R.~Bochicchio,
\newblock Bond orders and their relationships with cumulant and unpaired
  electron densities,
\newblock J. Phys. Chem. A {\bf 107}, 127 (2003).

\bibitem{Goddard:corr_chem_bond:1998}
T.~Yamasaki and W.~A. Goddard,
\newblock Correlation analysis of chemical bonds,
\newblock J. Phys. Chem. A {\bf 102}, 2919 (1998).

\bibitem{Luzanov:bond_indices:2005}
A.~V. Luzanov and O.~V. Prezhdo,
\newblock Irreducible charge density matrices for analysis of many-electron
  wave functions,
\newblock Int. J. Quant. Chem. {\bf 102}, 582 (2005).

\bibitem{Mayer:bond_order:2007}
I.~Mayer,
\newblock Bond order and valence indices: A personal account,
\newblock J. Comput. Chem. {\bf 28}, 204 (2007).

\bibitem{Davidson:local_spin:2001}
A.~E. Clark and E.~R. Davidson,
\newblock Local spin,
\newblock J. Chem. Phys. {\bf 115}, 7382 (2001).

\bibitem{Davidson:local_spin:2002}
E.~R. Davidson and A.~E. Clark,
\newblock Local spin {II},
\newblock Mol. Phys. {\bf 100}, 373 (2002).

\bibitem{Davidson:MolMagnets:2002}
E.~R. Davidson and A.~E. Clark,
\newblock Model molecular magnets,
\newblock J. Phys. Chem. A {\bf 106}, 7456 (2002).

\bibitem{Hess:local_spin:2005}
C.~Herrmann, M.~Reiher, and B.~A. Hess,
\newblock Comparative analysis of local spin definitions,
\newblock J. Chem. Phys. {\bf 122}, 034102 (2005).

\bibitem{Luzanov:SpinCorr:15}
A.~V. Luzanov, D.~Casanova, X.~Feng, and A.~I. Krylov,
\newblock Quantifying charge resonance and multiexciton character in coupled
  chromophores by charge and spin cumulant analysis,
\newblock J. Chem. Phys. {\bf 142}, 224104 (2015).

\bibitem{GTHBasis}
J.~VandeVondele and J.~Hutter,
\newblock Gaussian basis sets for accurate calculations on molecular systems in
  gas and condensed phases,
\newblock J. Chem. Phys. {\bf 127}, 114105 (2007).

\bibitem{GTHPseudo}
S.~Goedecker, M.~Teter, and J.~Hutter,
\newblock Separable dual-space {G}aussian pseudopotentials,
\newblock Phys. Rev. B {\bf 54}, 1703 (1996).

\bibitem{RI_auxbasis}
C.~H{\"a}ttig,
\newblock Optimization of auxiliary basis sets for {RI-MP2} and {RI-CC2}
  calculations: Core--valence and quintuple-$\zeta$ basis sets for {H} to {A}r
  and {QZVPP} basis sets for {L}i to {K}r,
\newblock Phys. Chem. Chem. Phys. {\bf 7}, 59 (2005).

\bibitem{Morosin:NiO:exchange_striction:1971}
L.~C. Bartel and B.~Morosin,
\newblock Exchange striction in {NiO},
\newblock Phys. Rev. B {\bf 3}, 1039 (1971).

\bibitem{Takeuchi:CoO:1979}
S.~Sasaki, K.~Fujino, and Y.~Takeuchi,
\newblock X-ray determination of electron-density distributions in oxides,
  {M}g{O}, {M}n{O}, {C}o{O}, and {N}i{O}, and atomic scattering factors of
  their constituent atoms,
\newblock Proc. Jpn. Acad. Ser. B {\bf 55}, 43 (1979).

\bibitem{Crisan:FeO:2011}
O.~Crisan and A.D. Crisan,
\newblock Phase transformation and exchange bias effects in mechanically
  alloyed {F}e/magnetite powders,
\newblock J. Alloys Compd. {\bf 509}, 6522 (2011).

\bibitem{Pokhilko:algs:2022}
P.~Pokhilko, C.-N. Yeh, and D.~Zgid,
\newblock Iterative subspace algorithms for finite-temperature solution of
  {D}yson equation,
\newblock J. Chem. Phys. {\bf 156}, 094101 (2022).

\bibitem{Monkhorst:Pack:k-grid:1976}
H.~J. Monkhorst and J.~D. Pack,
\newblock Special points for {B}rillouin-zone integrations,
\newblock Phys. Rev. B {\bf 13}, 5188 (1976).

\bibitem{EwaldProbeCharge}
Joachim Paier, Robin Hirschl, Martijn Marsman, and Georg Kresse,
\newblock The {P}erdew--{B}urke--{E}rnzerhof exchange-correlation functional
  applied to the {G2-1} test set using a plane-wave basis set,
\newblock J. Chem. Phys. {\bf 122}, 234102 (2005).

\bibitem{CoulombSingular}
Ravishankar Sundararaman and T.~A. Arias,
\newblock Regularization of the {C}oulomb singularity in exact exchange by
  {W}igner-{S}eitz truncated interactions: Towards chemical accuracy in
  nontrivial systems,
\newblock Phys. Rev. B {\bf 87}, 165122 (2013).

\bibitem{Yoshimi:IR:2017}
H.~Shinaoka, J.~Otsuki, M.~Ohzeki, and K.~Yoshimi,
\newblock Compressing {G}reen's function using intermediate representation
  between imaginary-time and real-frequency domains,
\newblock Phys. Rev. B {\bf 96}, 035147 (2017).

\bibitem{PYSCF}
Q.~Sun, T.~C. Berkelbach, N.~S. Blunt, G.~H. Booth, S.~Guo, Z.~Li, J.~Liu,
  J.~D. McClain, E.~R. Sayfutyarova, S.~Sharma, S.~Wouters, and G.~K. Chan,
\newblock Pyscf: the python-based simulations of chemistry framework,
\newblock Wiley Interdiscip. Rev.: Comput. Mol. Sci. {\bf 8}, e1340 (2017).

\bibitem{Rusakov16}
A.~A. Rusakov and D.~Zgid,
\newblock Self-consistent second-order {G}reen’s function perturbation theory
  for periodic systems,
\newblock J. Chem. Phys. {\bf 144}, 054106 (2016).

\bibitem{Yeh:X2C:GW:2022}
C.-N. Yeh, A.~Shee, Q.~Sun, E.~Gull, and D.~Zgid,
\newblock Relativistic self-consistent $gw$: Exact two-component formalism with
  one-electron approximation for solids,
\newblock Phys. Rev. B {\bf 106}, 085121 (2022).

\bibitem{Gabedit:2011}
A.~Allouche,
\newblock Gabedit-a graphical user interface for computational chemistry
  softwares,
\newblock J. Comput. Chem. {\bf 32}, 174 (2011).

\bibitem{povray}
Persistence of vision {(TM)} raytracer.

\bibitem{ase-paper}
A.~H. Larsen, J.~J. Mortensen, J.~Blomqvist, I.~E. Castelli, R.~Christensen,
  M.~Dułak, J.~Friis, M.~N. Groves, B.~Hammer, C.~Hargus, E.~D. Hermes, P.~C.
  Jennings, P.~B. Jensen, J.~Kermode, J.~R. Kitchin, E.~L. Kolsbjerg, J.~Kubal,
  K.~Kaasbjerg, S.~Lysgaard, J.~B. Maronsson, T.~Maxson, T.~Olsen, L.~Pastewka,
  A.~Peterson, C.~Rostgaard, J.~Schiøtz, O.~Schütt, M.~Strange, K.~S.
  Thygesen, T.~Vegge, L.~Vilhelmsen, M.~Walter, Z.~Zeng, and K.~W. Jacobsen,
\newblock The atomic simulation environment---a {P}ython library for working
  with atoms,
\newblock J. Phys.: Condens. Matter {\bf 29}, 273002 (2017).

\bibitem{Curtarolo:kpath:symmetry:2010}
W.~Setyawan and S.~Curtarolo,
\newblock High-throughput electronic band structure calculations: Challenges
  and tools,
\newblock Comput. Mater. Sci. {\bf 49}, 299 (2010).

\bibitem{Majumdar:NiO:MnO:DFT:J:2011}
T.~Archer, C.~D. Pemmaraju, S.~Sanvito, C.~Franchini, J.~He, A.~Filippetti,
  P.~Delugas, D.~Puggioni, V.~Fiorentini, R.~Tiwari, and P.~Majumdar,
\newblock Exchange interactions and magnetic phases of transition metal oxides:
  Benchmarking advanced ab initio methods,
\newblock Phys. Rev. B {\bf 84}, 115114 (2011).

\bibitem{Goodenough:direct_exchange:1960}
J.~B. Goodenough,
\newblock Direct cation--cation interactions in several oxides,
\newblock Phys. Rev. {\bf 117}, 1442 (1960).

\bibitem{Kanamori:exchange_mechanisms:1959}
J.~Kanamori,
\newblock Superexchange interaction and symmetry properties of electron
  orbitals,
\newblock Journal of Physics and Chemistry of Solids {\bf 10}, 87 (1959).

\bibitem{GK:rules:summary}
Y.~R. Pei,
\newblock Philip {A}nderson’s superexchange model,
\newblock
  \url{https://web.archive.org/web/20230323080536/https://courses.physics.ucsd.edu/2017/Fall/physics211a/specialtopic/1964.pdf},
  2017.

\bibitem{Halpern:superexchange:1966}
V.~Halpern and D.~Gabor,
\newblock A generalized mechanism for superexchange,
\newblock Proceedings of the Royal Society of London. Series A. Mathematical
  and Physical Sciences {\bf 291}, 113 (1966).

\bibitem{Chen:rep:space_groups:1985}
J.-Q. Chen, M.-J. Gao, and G.-Q. Ma,
\newblock The representation group and its application to space groups,
\newblock Rev. Mod. Phys. {\bf 57}, 211 (1985).

\bibitem{Kananenka15}
Alexei~A. Kananenka, Emanuel Gull, and Dominika Zgid,
\newblock Systematically improvable multiscale solver for correlated electron
  systems,
\newblock Phys. Rev. B {\bf 91}, 121111(R) (2015).

\bibitem{Tran_GW_SEET}
T.~N. Lan, A.~Shee, J.~Li, E.~Gull, and D.~Zgid,
\newblock Testing self-energy embedding theory in combination with {GW},
\newblock Phys. Rev. B {\bf 96}, 155106 (2017).

\bibitem{Tran_jcp_2015}
T.~N. Lan, A.~A. Kananenka, and D.~Zgid,
\newblock Communication: Towards ab initio self-energy embedding theory in
  quantum chemistry,
\newblock J. Chem. Phys. {\bf 143}, 241102 (2015).

\bibitem{Tran16}
T.~N.~Lan, A.~A. Kananenka, and D.~Zgid,
\newblock Rigorous ab initio quantum embedding for quantum chemistry using
  green's function theory: Screened interaction, nonlocal self-energy
  relaxation, orbital basis, and chemical accuracy,
\newblock J. Chem. Theory Comput. {\bf 12}, 4856 (2016).

\bibitem{Tran_useet}
L.~N. Tran, S.~Iskakov, and D.~Zgid,
\newblock Spin-unrestricted self-energy embedding theory,
\newblock J. Phys. Chem. Lett. {\bf 9}, 4444 (2018).

\bibitem{Zgid17}
Dominika Zgid and Emanuel Gull,
\newblock Finite temperature quantum embedding theories for correlated systems,
\newblock New Journal of Physics {\bf 19}, 023047 (2017).

\bibitem{Rusakov:SEET:2019}
A.~A. Rusakov, S.~Iskakov, L.~N. Tran, and D.~Zgid,
\newblock Self-energy embedding theory ({SEET}) for periodic systems,
\newblock J. Chem. Theory Comput. {\bf 15}, 229 (2019).

\bibitem{Yeh:perovskite:2021}
C.-N. Yeh, S.~Iskakov, D.~Zgid, and E.~Gull,
\newblock Electron correlations in the cubic paramagnetic perovskite
  $\mathrm{Sr}(\mathrm{V},\mathrm{Mn}){\mathrm{o}}_{3}$: Results from fully
  self-consistent self-energy embedding calculations,
\newblock Phys. Rev. B {\bf 103}, 195149 (2021).

\bibitem{deGraaf:CaMnO3:charge_transfer:2007}
A.~Sadoc, R.~Broer, and C.~de~Graaf,
\newblock Role of charge transfer configurations in {L}a{M}n{O}3, {C}a{M}n{O}3,
  and {C}a{F}e{O}3,
\newblock J. Chem. Phys. {\bf 126}, 134709 (2007).

\bibitem{HSE:2003}
J.~Heyd, G.~E. Scuseria, and M.~Ernzerhof,
\newblock Hybrid functionals based on a screened {C}oulomb potential,
\newblock J. Chem. Phys. {\bf 118}, 8207 (2003).

\end{thebibliography}
\end{document}